\documentclass[11pt,a4paper]{article}

\usepackage[margin=1in]{geometry}
\usepackage{caption}

\usepackage{caption}

\DeclareCaptionFont{tenpt}{%
  \fontsize{10pt}{11pt}\selectfont
}

\usepackage[utf8]{inputenc}
\usepackage[T1]{fontenc}
\usepackage[english]{babel}
\usepackage{lmodern}
\usepackage{microtype}
\usepackage{authblk}
\usepackage{pdflscape}
\usepackage{array}
\usepackage{amsmath,amssymb,amsthm}
\usepackage{bm}
\usepackage{upgreek}

\usepackage{graphicx}
\usepackage{booktabs}
\usepackage{tabularx}
\usepackage{multirow}
\usepackage{dcolumn}
\usepackage{subcaption}

\usepackage{xcolor}               
\usepackage{soul}                 

\usepackage{float}

\usepackage{titlesec}
\titleformat{\section}{\normalfont\large\bfseries}{\thesection}{1em}{}
\titleformat{\subsection}{\normalfont\normalsize\bfseries}{\thesubsection}{1em}{}
\titleformat{\subsubsection}{\normalfont\normalsize\bfseries}{\thesubsubsection}{1em}{}

\renewcommand{\thesection}{\arabic{section}}
\renewcommand{\thesubsection}{\thesection.\arabic{subsection}}
\renewcommand{\thesubsubsection}{\thesubsection.\arabic{subsubsection}}

\usepackage{makecell}

\usepackage[colorlinks=true,linkcolor=blue,citecolor=blue,urlcolor=blue]{hyperref}

\theoremstyle{plain}
\newtheorem{thm}{Theorem}

\newtheorem{prop}[thm]{Proposition}
\newtheorem{cor}[thm]{Corollary}
\newtheorem{rem}[thm]{Remark}

\begin{document}

\title{An Infinite Two-Parameter Family of Octagonal Tilings}

\author[1]{April Lynne D. Say-awen}
\author[2]{Sam Coates}

\affil[1]{\small{Mathematics and Statistics Department, De La Salle University, Malate, Manila, 1004, Philippines}}
\affil[2]{\small{Surface Science Research Centre, Department of Physics, University of Liverpool, Liverpool L69 3BX, United Kingdom}}

\date{\small{\today}}

\maketitle
\abstract{\noindent We construct an infinite two-parameter family of substitution rules with inflation factors $\delta_{(m,n)}=m+n(1+\sqrt{2})$ for integers $m\geq0$ and $n\geq1$, excluding pairs in which $m$ is even and $n$ is odd. Extending four examples from our previous work ~\cite{saycoa2026}, we show that each substitution admits a tiling with $8$-fold rotational symmetry and that the associated tilings are repetitive and have finite local complexity. We also classify the inflation factors according to their PV and unimodular properties and illustrate selected cases through perpendicular-space projections and Fourier spectra.}

\section{Introduction and motivation}\label{sec:intro}
Tilings with rotational symmetries forbidden by the crystallographic restriction provide important models of planar aperiodic order. Among these, $8$-fold rotational symmetry is one of the most extensively studied, owing to its close connection with octagonal quasicrystals and higher-dimensional lattice constructions. A fundamental example is the Ammann--Beenker tiling \cite{ammann1992aperiodic,grunbaumshephard1987}, which can be constructed by a substitution rule on a square and a rhombus with acute angle $\pi/4$, using the silver mean $\mu=1+\sqrt{2}$ as its inflation factor. Watanabe et al.~\cite{watanabe1987} constructed a related octagonal tiling with inflation factor $2+\sqrt{2}$. More recently, in~\cite{saycoa2026}, we introduced four examples based on a common set of prototiles, with inflation factors $2+\sqrt{2}$, $3+2\sqrt{2}$, $4+2\sqrt{2}$, and $5+2\sqrt{2}$.

These inflation factors belong to the quadratic integer ring $\mathbb{Z}[\sqrt{2}]$ and have the common form $\delta_{(m,n)}=m+n\mu$. In this parametrization, the Ammann--Beenker tiling corresponds to $(m,n)=(0,1)$, while the Watanabe--Ito--Soma tiling and the first example from~\cite{saycoa2026} have parameter pair $(1,1)$. The remaining examples from~\cite{saycoa2026} correspond to $(1,2)$, $(2,2)$, and $(3,2)$. In the present work, we place these examples within a unified two-parameter framework. More precisely, for every pair of integers $m\geq0$ and $n\geq1$, except when $m$ is even and $n$ is odd, we construct a substitution rule $\sigma_{(m,n)}$ with inflation factor $\delta_{(m,n)}$ that admits a tiling with $8$-fold rotational symmetry. The excluded parity case is not treated because it is incompatible with the symmetric edge dissections imposed in our construction; this does not rule out the existence of substitution rules obtained from different edge dissections. The resulting family extends the Ammann--Beenker construction from the silver-mean case $\delta_{(0,1)}=\mu$ to an infinite collection of inflation factors in $\mathbb{Z}[\sqrt{2}]$.

The direct physical motivation for our generalisation comes from recent work on binary hard spheres. Fayen et al.~\cite{fayen2023self} showed that a mixture of large and small hard spheres resting on a flat substrate self-assembles into two families of random-tiling quasicrystalline phases, one dodecagonal and one octagonal. The authors noted that, at the time, there was no known tiling that could characterise the octagonal phases. Importantly, these phases are composed of random tilings whose tile compositions vary continuously with the radius of the small spheres, $x_s$. The tiling family introduced here thus provides an infinite collection of perfectly ordered, $8$-fold symmetric reference structures whose compositions can be tuned through $(m,n)$, or, conversely, by $x_s$: the observed phases may therefore be modelled as randomised versions of tilings in the family.

The paper is organised as follows. Section~\ref{sec:basics} collects the definitions and standard results on tilings and substitutions used throughout. Section~\ref{sec:construction} describes the edge dissections and the decompositions of the inflated prototiles, and establishes Theorem~\ref{thm:maintheorem}: for every admissible pair $(m,n)$, the substitution $\sigma_{(m,n)}$ admits a tiling $\mathcal{T}_{(m,n)}$ with $8$-fold rotational symmetry. This tiling is invariant under $D_8$ when $m+n$ is even, and under $C_8$, without reflection symmetry, when $m+n$ is odd. In Section~\ref{sec:properties} we compute the substitution matrices, prove primitivity, and derive the prototile frequencies and area fractions. Together with simplicity, these results yield finite local complexity and repetitivity of all associated tilings. Section~\ref{sec:algebra} classifies the algebraic properties of the inflation factors: we show that $\delta_{(m,n)}$ is a Pisot--Vijayaraghavan number exactly when $m=\lfloor n(\sqrt{2}-1)\rfloor$ or $m=\lfloor n(\sqrt{2}-1)\rfloor+1$. We also show that the substitution is unimodular exactly when $\delta_{(m,n)}=\mu^{2r}$, that is, when $(m,n)=(P_{2r-1},P_{2r})$ for consecutive Pell numbers. Section~\ref{sec:perp} presents perpendicular-space images and Fourier spectra for $\mathcal{T}_{(1,1)}$, $\mathcal{T}_{(1,2)}$ and $\mathcal{T}_{(5,12)}$. Section~\ref{sec:comparison} relates the family to the octagonal hard-sphere phases, and Section~\ref{sec:conclusion} concludes with an outlook.

\section{Preliminaries}\label{sec:basics}
\noindent 
A \textit{tile} is a subset of $\mathbb{R}^2$ homeomorphic to the closed unit disk. Throughout this paper, all tiles are assumed to be polygons. The \textit{support} of a collection $\mathcal{A}$ of tiles is $\mbox{supp}(\mathcal{A})=\bigcup_{T\in\mathcal{A}}T$. A \textit{patch} is a finite collection of tiles with pairwise disjoint interiors and simply connected support.
 
 A \textit{tiling} of $\mathbb{R}^2$ is a countable collection $\mathcal{T}$ of tiles with pairwise disjoint interiors such that $\mbox{supp}(\mathcal{T})=\mathbb{R}^2$. For $r>0$, an \textit{$r$-patch} of $\mathcal{T}$ is a patch whose support is contained in a ball of radius $r$.

Let $\mbox{Isom}(\mathbb{R}^2)$ denote the group of Euclidean isometries of $\mathbb{R}^2$. Every $g\in\mbox{Isom}(\mathbb{R}^2)$ can be written uniquely as $g(x)=Rx+t,$ where $t\in\mathbb{R}^2$ and $R\in O(2)$, the group of rotations and reflections fixing the origin. For a tile $T$ and a collection 
$\mathcal{A}$ of tiles, define
$$
g(T)=RT+t=\{Rx+t\mid x\in T\},
\qquad
g(\mathcal{A})=\{g(T)\mid T\in\mathcal{A}\}.
$$
If $R$ is the trivial rotation, we write $g(T)=T+t$ and $g(\mathcal{A})=\mathcal{A}+t$, and call these \textit{translated copies}. If $R$ is a nontrivial rotation or a reflection, we call them \textit{rotated copies} or \textit{reflected copies}, respectively.

Two tiles $T$ and $T'$ are \textit{congruent} if $T'=g(T)$ for some $g\in\mbox{Isom}(\mathbb{R}^2)$. In this case, $T$ and $T'$ belong to the same congruence class. We assume that every tiling $\mathcal{T}$ under consideration has finitely many congruence classes of tiles. A choice of one representative from each class forms a \textit{prototile set} $\mathcal{S}=\{T_1,\ldots,T_m\}.$ Thus, every $T\in\mathcal{T}$ can be expressed as $T=RT_i+t$ for some $i\in\{1,\ldots,m\}$, $R\in O(2)$, and $t\in\mathbb{R}^2$. Define the set of orientations occurring in $\mathcal{T}$ by
$$
\mathcal{O}(\mathcal{T})=\left\{ R\in O(2) \mid T=RT_i+t \text{ for some } T\in\mathcal{T},\ i\in\{1,\ldots,m\},\ t\in\mathbb{R}^2 \right\}.
$$

The tiles of $\mathcal{T}$ occur in \textit{finitely many orientations} if $\mathcal{O}(\mathcal{T})$ is finite. Since the tiles are polygons, this is equivalent to requiring that $\mathcal{T}$ have only finitely many translation classes of tiles, where two tiles belong to the same translation class if one is a translated copy of the other.

A tiling is \textit{edge-to-edge} if the intersection of any two distinct tiles is either empty, a common vertex, or a full edge of both tiles. An edge-to-edge tiling is called \textit{simple} if its tiles occur in finitely many orientations.

The notions of congruence and translation equivalence extend naturally to patches through the action defined above. A tiling $\mathcal{T}$ has \textit{finite local complexity} (FLC) if, for every $r>0$, it contains only finitely many translation classes of $r$-patches. Finally, a tiling $\mathcal{T}$ is \textit{repetitive} if, for every patch $\mathcal{P}\subseteq\mathcal{T}$, there exists $\rho>0$ such that every ball of radius $\rho$ contains the support of a translated copy of $\mathcal{P}$ occurring in $\mathcal{T}$.

\begin{rem}\label{rem:flc}
A tiling has FLC if and only if it has finitely many translation classes of tiles and finitely many translation classes of two-tile patches~\cite{perrymansadun2025}. Consequently, every simple tiling has FLC.
\end{rem}

The \textit{symmetry group} of a tiling $\mathcal{T}$ is defined by $\mbox{Sym}(\mathcal{T})=\{g\in \mbox{Isom}(\mathbb{R}^2)\mid g(\mathcal{T})=\mathcal{T}\}$. The tiling $\mathcal{T}$ is \textit{periodic} if $\mbox{Sym}(\mathcal{T})$ contains a nontrivial translation; equivalently, if $\mathcal{T}+t=\mathcal{T}$ for some nonzero $t\in\mathbb{R}^2$. It has \textit{rotational symmetry of order $q$}, or \textit{$q$-fold rotational symmetry}, if $\mbox{Sym}(\mathcal{T})$ contains a rotation of order~$q$.

The main objective of this work is to construct repetitive tilings with finite local complexity and $8$-fold rotational symmetry. We obtain these tilings using substitution rules.

Let $\mathcal{S}=\{T_1,\ldots,T_m\}$ be a finite set of pairwise noncongruent tiles, and let $\delta>1$. A \textit{substitution rule} $\sigma$ on $\mathcal{S}$ with \textit{inflation factor} $\delta$ assigns to each $T_i\in\mathcal{S}$ a patch $\sigma(T_i)$ of tiles, each congruent to an element of $\mathcal{S}$. In this work, we require $\delta T_i\subseteq\mbox{supp}(\sigma(T_i))$. Every tile of $\sigma(T_i)$ not contained in $\delta T_i$ crosses the boundary of $\delta T_i$ and is bisected by it. Thus, $\delta T_i$ is the inflated region associated with $T_i$, although $\sigma(T_i)$ may extend beyond this region. Accordingly, $\sigma$ may be viewed as an inflate-and-subdivide rule in which $T_i$ is inflated by $\delta$ and the resulting region $\delta T_i$ is decomposed into whole tiles and halves of the boundary tiles.

The rule is required to extend consistently to congruent copies and finite collections of tiles and to remain well-defined under iteration. If $T=RT_i+t$ is a congruent copy of $T_i$, where $R\in O(2)$ and $t\in\mathbb{R}^2$, then $\sigma(T)=R(\sigma(T_i))+\delta t$. For a patch $\mathcal{A}$ of tiles to which the substitution is applied, define $\sigma(\mathcal{A})=\bigcup_{T\in\mathcal{A}}\sigma(T)$. This extension must satisfy the following compatibility condition: for any $T, T' \in \mathcal{A}$, if a tile in $\sigma(T)$ and a tile in $\sigma(T')$ overlap in their interiors, then they coincide and occur only once in $\sigma(\mathcal{A})$; partial overlaps are not permitted. For $k\geq1$, define $\sigma^{k+1}(T_i)=\sigma(\sigma^k(T_i))$. The collection $\sigma^k(T_i)$ is called the \textit{$k$th-order supertile} of $T_i$.

A \textit{legal patch} of $\sigma$ is any patch congruent to a subpatch of $\sigma^k(T_i)$ for some $T_i\in\mathcal{S}$ and some $k\geq0$, where $\sigma^0(T_i)=T_i$. A tiling $\mathcal{T}$ is called a \textit{substitution tiling} associated with $\sigma$ if every patch of $\mathcal{T}$ is legal. The set $\mathcal{S}$ is called the \textit{prototile set} of $\sigma$, and its elements are called the \textit{prototiles} of $\sigma$. If every congruence class represented in $\mathcal{S}$ occurs in $\mathcal{T}$, then $\mathcal{S}$ is also a prototile set for $\mathcal{T}$ in the sense defined earlier. Otherwise, a prototile set for $\mathcal{T}$ is given by the proper subset of $\mathcal{S}$ consisting of those prototiles whose congruence classes occur in $\mathcal{T}$.

The collection of all substitution tilings associated with $\sigma$ is called the \textit{hull} of $\sigma$, denoted by $\mathbb{X}_{\sigma}$.

A basic method for obtaining a tiling from $\sigma$ is to choose a legal patch $\mathcal{P}$ for which there exists an integer $k\geq1$ such that $\mathcal{P}\subseteq\sigma^k(\mathcal{P})$, with the origin lying in the interior of $\mbox{supp}(\mathcal{P})$. The patches $\mathcal{P},\sigma^k(\mathcal{P}),\sigma^{2k}(\mathcal{P}),\ldots$ then form a nested sequence. Since $\delta^{sk}\mbox{supp}(\mathcal{P})\subseteq \mbox{supp}(\sigma^{sk}(\mathcal{P}))$, their supports expand to cover $\mathbb{R}^2$. Hence, their union $\mathcal{T}=\bigcup_{s\geq0}\sigma^{sk}(\mathcal{P})$, where $\sigma^0(\mathcal{P})=\mathcal{P}$, is a tiling in $\mathbb{X}_{\sigma}$. The patch $\mathcal{P}$ is called a \textit{seed} of $\mathcal{T}$, and the resulting tiling satisfies $\sigma^k(\mathcal{T})=\mathcal{T}$.

A desirable property of a substitution rule is primitivity. A substitution rule $\sigma$ is called \textit{primitive} if there exists an integer $k\geq1$ such that, for every $T_i\in\mathcal{S}$, the supertile $\sigma^k(T_i)$ contains a congruent copy of every prototile in $\mathcal{S}$. A standard result states that primitivity, together with finite local complexity, implies repetitivity; see \cite{Frank2008,FrettlohRichard2013,baakegrimm2013}.

\begin{rem}
\label{rem:repetitive}
A substitution tiling with FLC arising from a primitive substitution rule is repetitive.
\end{rem}

Primitivity of $\sigma$ can also be checked using its substitution matrix $M_\sigma$. The $(i,j)$-entry $a_{ij}$ of $M_\sigma$ records the number of copies congruent to $T_i$ in the decomposition of the inflated prototile $\delta T_j$. Each copy contained in $\delta T_j$ is counted with weight $1$, while each copy bisected by its boundary is counted with weight $1/2$. Thus, $\mbox{area}(\delta T_j)=\delta^2\mbox{area}(T_j)=\sum_{i=1}^{m}a_{ij}\mbox{area}(T_i)$.

By compatibility of the decompositions under iteration, the $(i,j)$-entry of $M_\sigma^k$ gives the number of copies congruent to $T_i$ in the decomposition of $\delta^kT_j$, with copies contained in $\delta^kT_j$ counted with weight $1$ and copies bisected by its boundary counted with weight $1/2$. Consequently, $\sigma$ is primitive if and only if $M_\sigma$ is primitive; that is, if there exists $k\in\mathbb{N}$ such that every entry of $M_\sigma^k$ is positive.

\newpage
\section{Construction}\label{sec:construction}

\begin{figure}[H]
		\begin{subfigure}{0.45\textwidth}
			\centering
			\includegraphics[width=.8\textwidth]{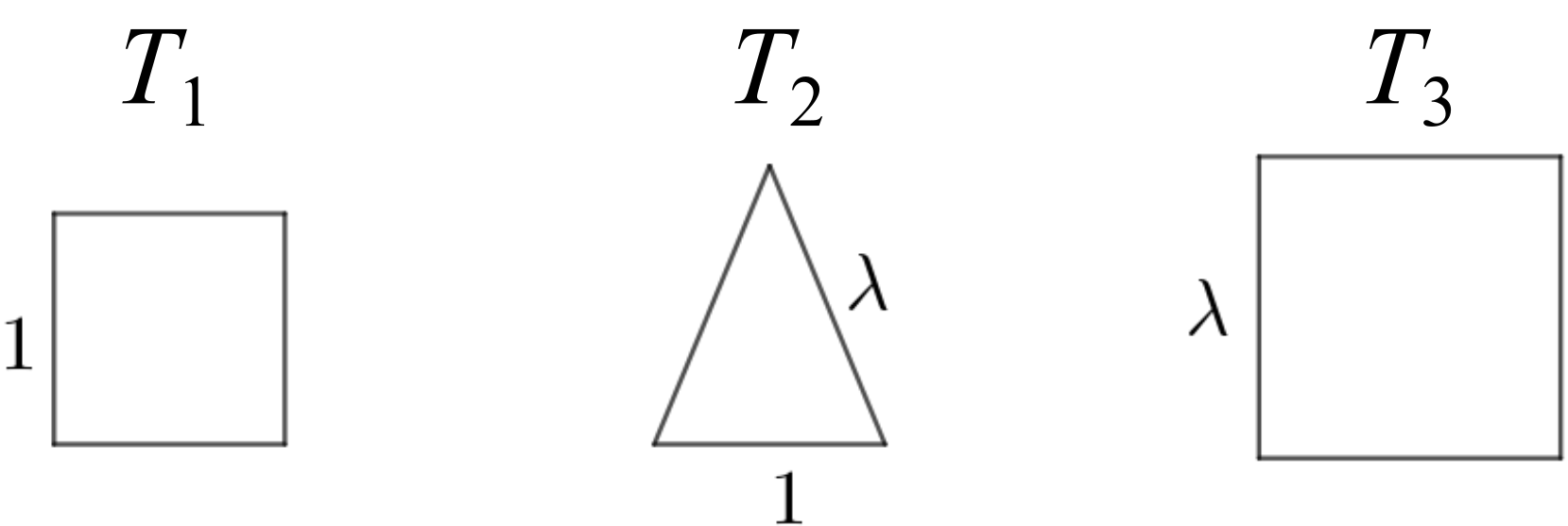}
			\caption{}
			\label{fig:prototiles}
		\end{subfigure}
		\hfill
		\begin{subfigure}{0.45\textwidth}
			\centering
			\includegraphics[width=.8\textwidth]{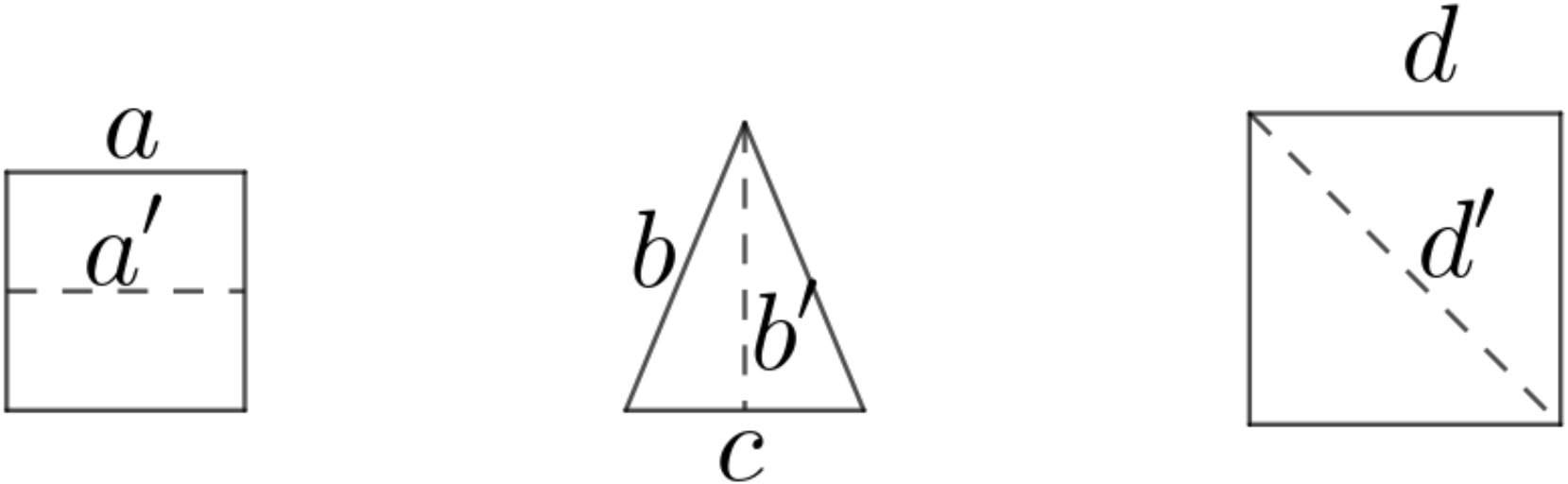}
			\caption{}
			\label{fig:edgelabels}
		\end{subfigure}
		
		\caption{(a) The prototiles; (b) edge labeling.}
		\label{fig:prototileslabel}
	\end{figure}

Our construction begins with the three polygons shown in Figure~\ref{fig:prototiles}: a unit square $T_1$, an isosceles triangle $T_2$ with apex angle $\pi/4$, base length $1$, and equal sides of length $\lambda=\sqrt{\frac{2+\sqrt{2}}{2}}$, and a square $T_3$ of side length $\lambda$.

Let $\mu=1+\sqrt{2}$ denote the silver mean. For integers $m,n$ with $m\geq 0$ and $n\geq 1$, we define the inflation factor $\delta_{(m,n)}=m+n\mu$ excluding the case in which $m$ is even and $n$ is odd.

The choice of inflation factors is dictated by the geometry of the prototiles. The square $T_2$ and triangle $T_3$ are precisely the prototiles used in the Ammann--Beenker tiling, whose inflation factor is the silver mean $\mu$. 
The addition of the smaller square $T_1$ leads to the inflation factor $\delta_{(1,1)}=1+\mu$ and, more generally, to the family $\delta_{(m,n)}=m+n\mu$. We restrict attention to symmetric edge dissections. As will become clear in Remark~\ref{rem:parity-restriction}, this symmetry condition excludes the case in which $m$ is even and $n$ is odd.

For each prototile edge $e$, we specify a dissection of $\delta_{(m,n)}e$ into tile edges and distinguished segments that bisect certain tiles into two congruent halves. As shown in Figure~\ref{fig:edgelabels}, the edges of $T_1$ are labeled $a$, the legs and base of $T_2$ are labeled $b$ and $c$, respectively, and the edges of $T_3$ are labeled $d$. The chosen bisecting segments of $T_1$, $T_2$, and $T_3$ are labeled $a'$, $b'$, and $d'$, respectively. Within each prototile, edges of the same length carry the same label, and all edges carrying a given label are assigned the same dissection. For any line segment $s$, we denote its length by $|s|$.

Note that every prototile edge has length either $1$ or $\lambda$. In terms of the labeled segments in Figure~\ref{fig:edgelabels}, the inflated unit edge length is

\begin{equation}
\delta_{(m,n)}=m+n\mu=m|a|+2n|b'|.
\label{eq:delta}
\end{equation}
where a segment labeled $a$ may be replaced by one labeled $a'$ or $c$, since $|a|=|a'|=|c|$. Similarly,

\begin{equation}
\delta_{(m,n)}\lambda
=m\lambda+n\lambda\mu
=(m+n)|b|+n|d'|,
\label{eq:deltalambda}
\end{equation}
where a segment labeled $b$ may be replaced by one labeled $d$, since $|b|=|d|$.

\begin{rem}[Symmetric edge dissections]
\label{rem:parity-restriction}
For simplicity, we require each edge dissection to be symmetric about the midpoint of the inflated edge. A dissection of an edge of length $\delta_{(m,n)}\lambda$, as in~\eqref{eq:deltalambda}, contains $m+n$ segments labeled $b$ or $d$, and $n$ segments labeled $d'$. In a symmetric arrangement, segments occur in reflected pairs, except possibly for a single segment centered at the midpoint. Hence, at least one of the integers $m+n$ and $n$ must be even. If $n$ is odd, then $m+n$ must be even, which forces $m$ to be odd. Therefore, the case in which $m$ is even and $n$ is odd is excluded from the present construction.
\end{rem}

The edge dissections, and hence the corresponding substitution rules, are obtained from \eqref{eq:delta} and \eqref{eq:deltalambda}. The construction splits into two cases according to the parity of $m+n$.
\newpage

\noindent \textbf{Case 1: $m+n$ is even}\\

\begin{figure}[H]
		\centering
		\includegraphics[scale=.9]{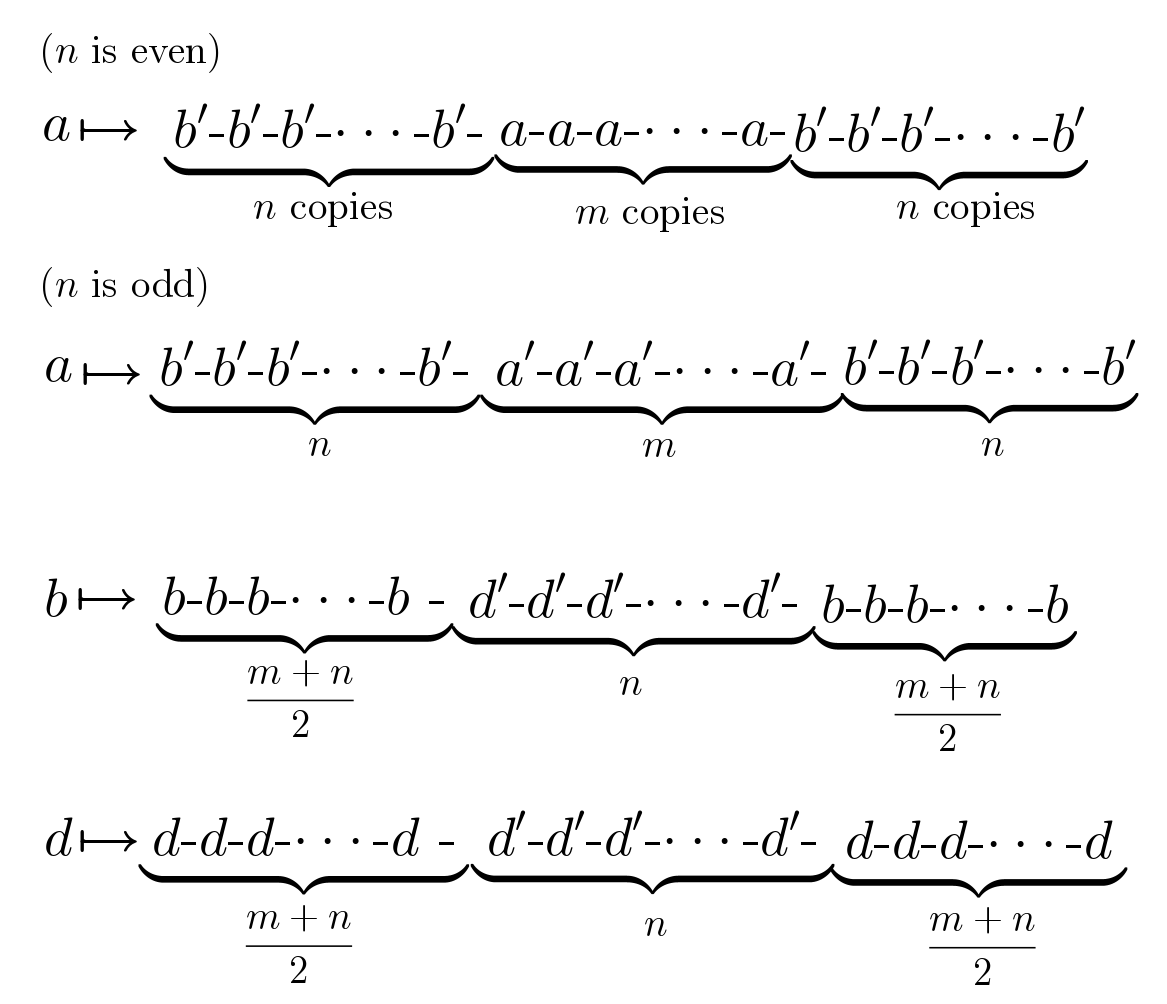}
\caption{Edge dissection rule for the case where $m+n$ is even; the dissection for $\delta_{(m,n)}c$ is the same as that for $\delta_{(m,n)}a$.}
		\label{fig:edgesubstitution}
	\end{figure}

\noindent Suppose first that $m+n$ is even. The corresponding edge dissections are shown in Figure~\ref{fig:edgesubstitution}. In this case, the images of $a$ and $c$ are identical. An occurrence of $b'$ in the image of $a$ indicates that a copy of $T_2$ crossing the boundary is bisected by the inflated edge  $\delta_{(m,n)}a$; occurrences of $a'$ and $d'$ have the analogous interpretation. The copies of $T_2$ are oriented so that the apices of the two copies at the ends coincide with the endpoints of $\delta_{(m,n)}a$, and they meet in pairs along their bases. When $n$ is odd, however, the two copies in the central pair are instead
adjacent to tiles congruent to $T_1$. The given edge dissections allow two tiles to meet along edges of the same
length.

Once the boundary of $\delta_{(m,n)}T_i$ is dissected according to Figure~\ref{fig:edgesubstitution}, we partition the inflated prototile into regions, each consisting of copies of a single prototile $T_j$. The edge dissection prescribes the positions of the tiles crossed by the boundary of $\delta_{(m,n)}T_i$. Each such tile occurs as a whole tile in the substitution patch, but only one of its two congruent halves lies within the inflated prototile; the other half extends into the adjacent inflated prototile. When counting the tiles in the decomposition of $\delta_{(m,n)}T_i$, each bisected boundary tile is therefore assigned weight $1/2$. Thus, a region containing $p$ whole copies of $T_j$ and $q$ bisected boundary copies is counted as $p+q/2$ copies of $T_j$.

The shapes of the regions are determined by the edge lengths and interior angles of the prototiles. For example, the scaled triangle $nT_2$ can be dissected into $n^2$ congruent copies of $T_2$ by subdividing each side into $n$ equal parts and drawing lines through the subdivision points parallel to the sides.

\begin{figure}[H]
		\centering
		\includegraphics[width=.75\linewidth]{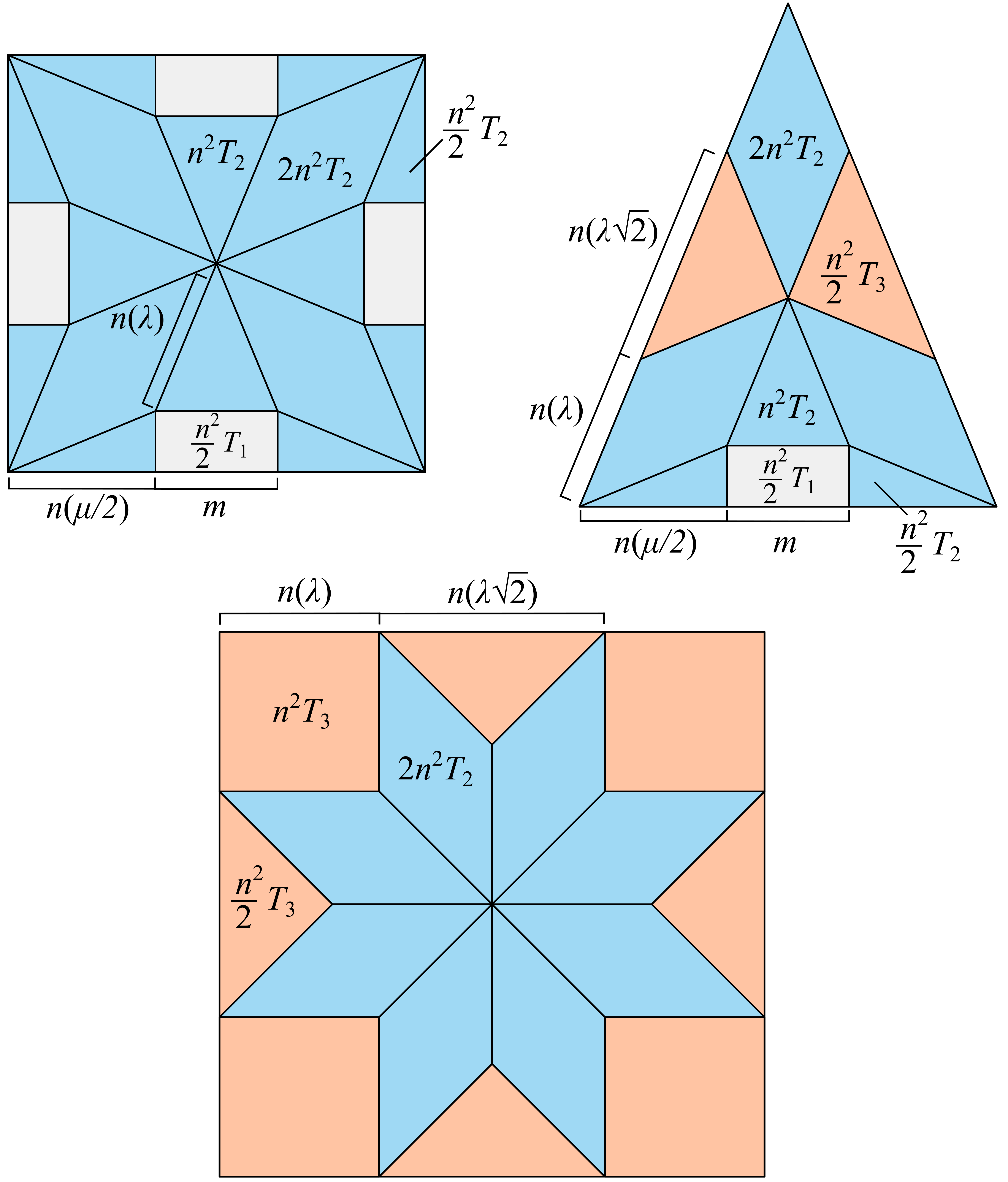}
		\caption{The dissections of $\delta_{(n,n)}T_i$ into regions, for $i\in\{1,2,3\}$. Each region can be further dissected into copies of a single prototile, with the number of tiles indicated in selected regions.}
		\label{fig:casem=n}
	\end{figure}

\begin{figure}[H]
		\centering
		\includegraphics[width=.75\linewidth]{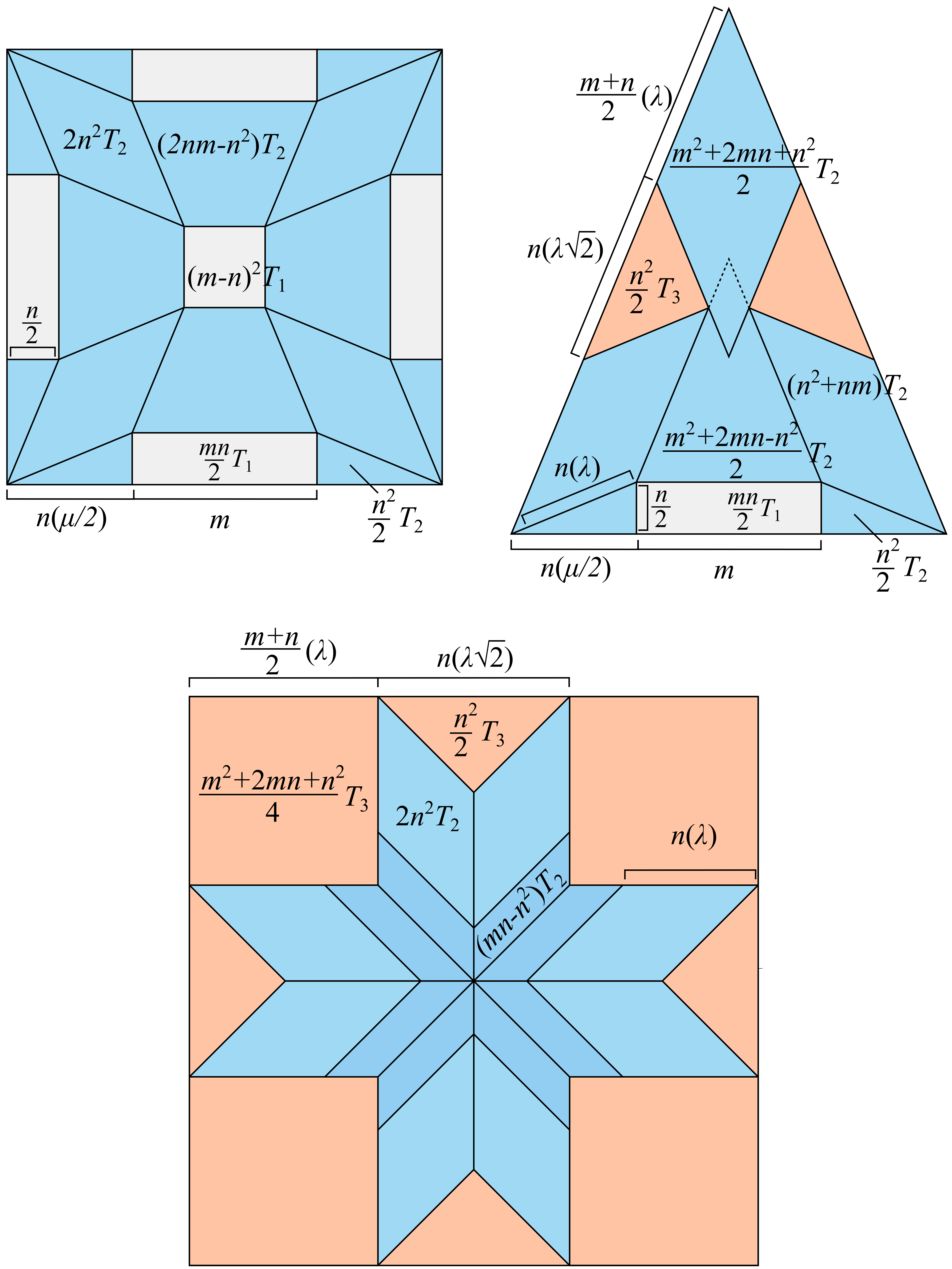}
		\caption{The dissection of $\delta_{(m,n)} T_i$, $i \in \{1,2,3\}, m>n.$}
		\label{fig:casemGreatern}
	\end{figure}

\begin{figure}[H]
		\centering
		\includegraphics[width=.75\linewidth]{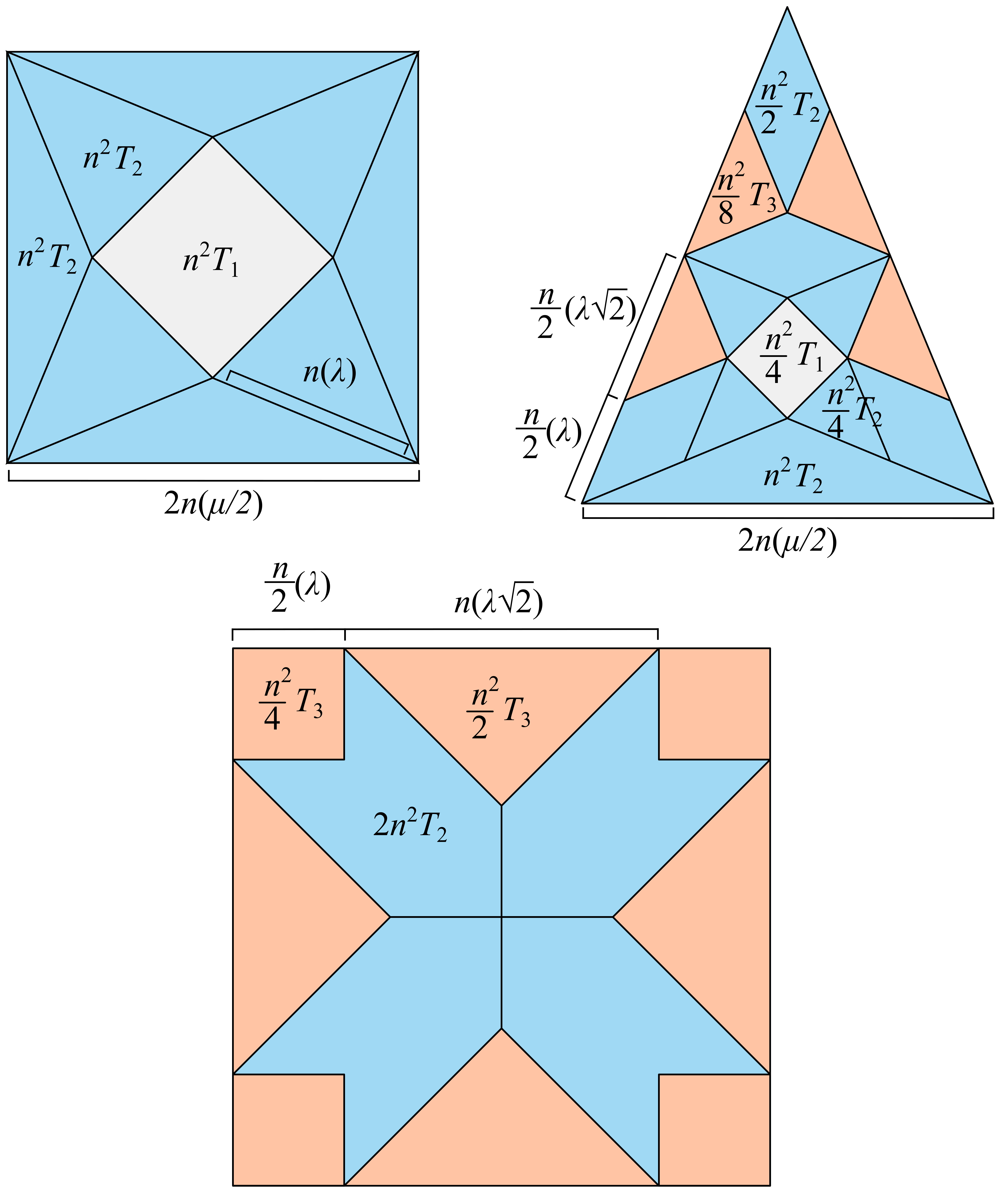}
		\caption{The dissection of $\delta_{(m,n)} T_i$, $i \in \{1,2,3\}, m=0.$}
		\label{fig:casem=0}
	\end{figure}

\begin{figure}[H]
		\centering
		\includegraphics[width=\linewidth]{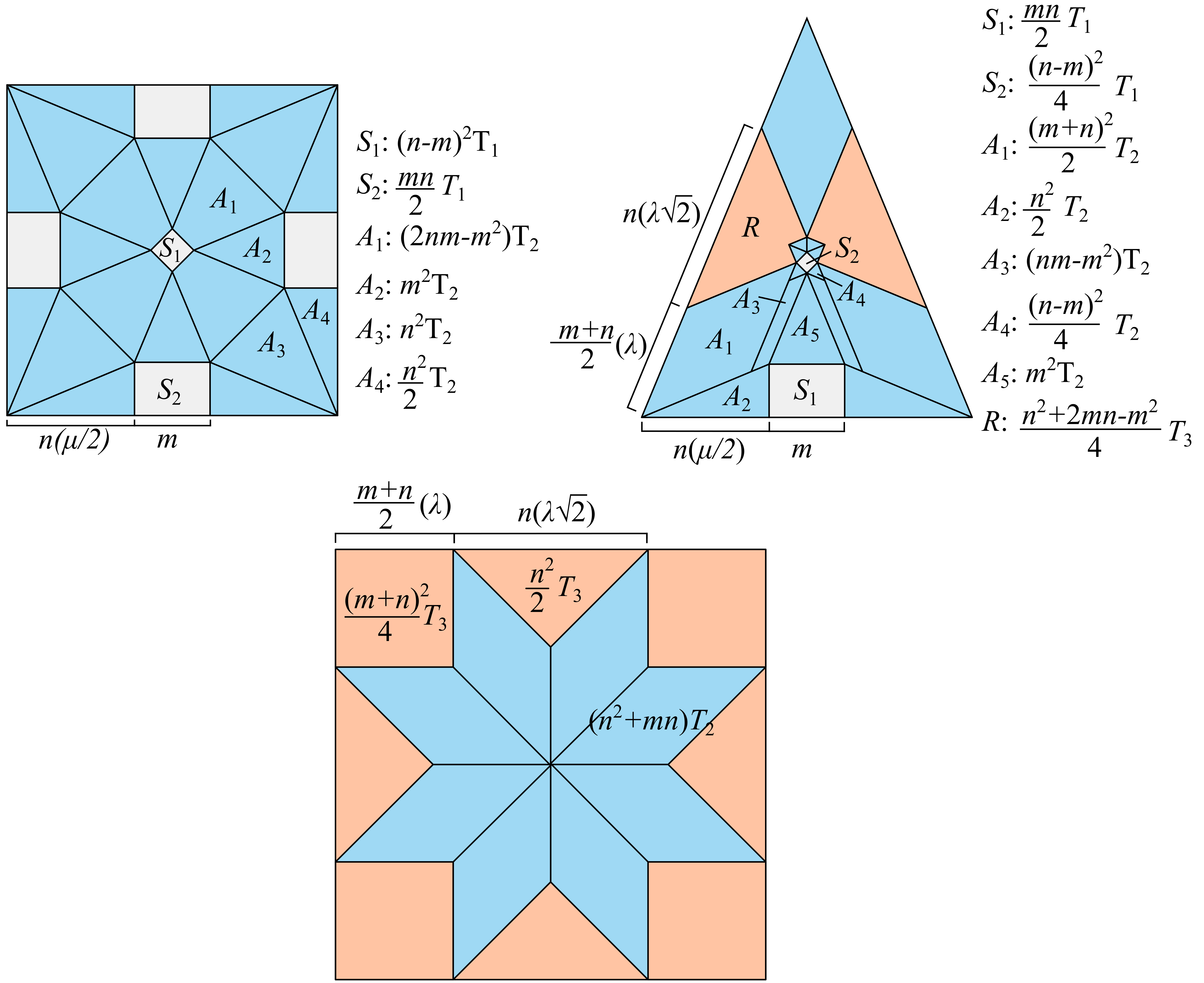}
		\caption{The dissection of $\delta_{(m,n)} T_i$, $i \in \{1,2,3\}, 0<m<n$.}
		\label{fig:case0mn}
	\end{figure}

Figures~\ref{fig:casem=n}--\ref{fig:case0mn} show the decompositions of the inflated prototiles for the cases $m=n$, $m>n$, $m=0$, and $0<m<n$, respectively. Each gray region is tiled by copies congruent to $T_1$, each blue region by copies congruent to $T_2$, and each orange region by copies congruent to $T_3$. The number of copies in each region, with bisected boundary tiles counted with weight $1/2$, is indicated within or beside the region, and the relevant edge lengths are labeled. Together with the prescribed edge dissections, these decompositions define the substitution rule $\sigma_{(m,n)}$.

In each case, a patch $\mathcal{H}$ consisting of eight copies of $T_2$ arranged with $8$-fold rotational symmetry lies at the center of $\sigma_{(m,n)}(T_3)$. We center $\mathcal{H}$ at the origin and apply $\sigma_{(m,n)}$ to obtain $\mathcal{H}\subseteq \sigma_{(m,n)}(\mathcal{H})$. Hence, the patches $\sigma_{(m,n)}^k(\mathcal{H})$ form a nested sequence. Their supports expand to cover $\mathbb{R}^2$, so $\mathcal{T}_{(m,n)}=\bigcup_{k\geq 0}\sigma_{(m,n)}^k(\mathcal{H})$ is a tiling corresponding to $\sigma_{(m,n)}$.

The substitution respects the symmetries of each prototile and therefore preserves the symmetries of $\mathcal{H}$. Consequently, $\mathcal{T}_{(m,n)}$ is invariant under $D_8$, the dihedral group of order $16$, and in particular has $8$-fold rotational symmetry. The construction is illustrated for $\sigma_{(0,2)}$ in Figures~\ref{fig:0,2substitution} and~\ref{fig:0,2seed}.

	\begin{figure}[H]
		\centering
		\includegraphics[width=\linewidth]{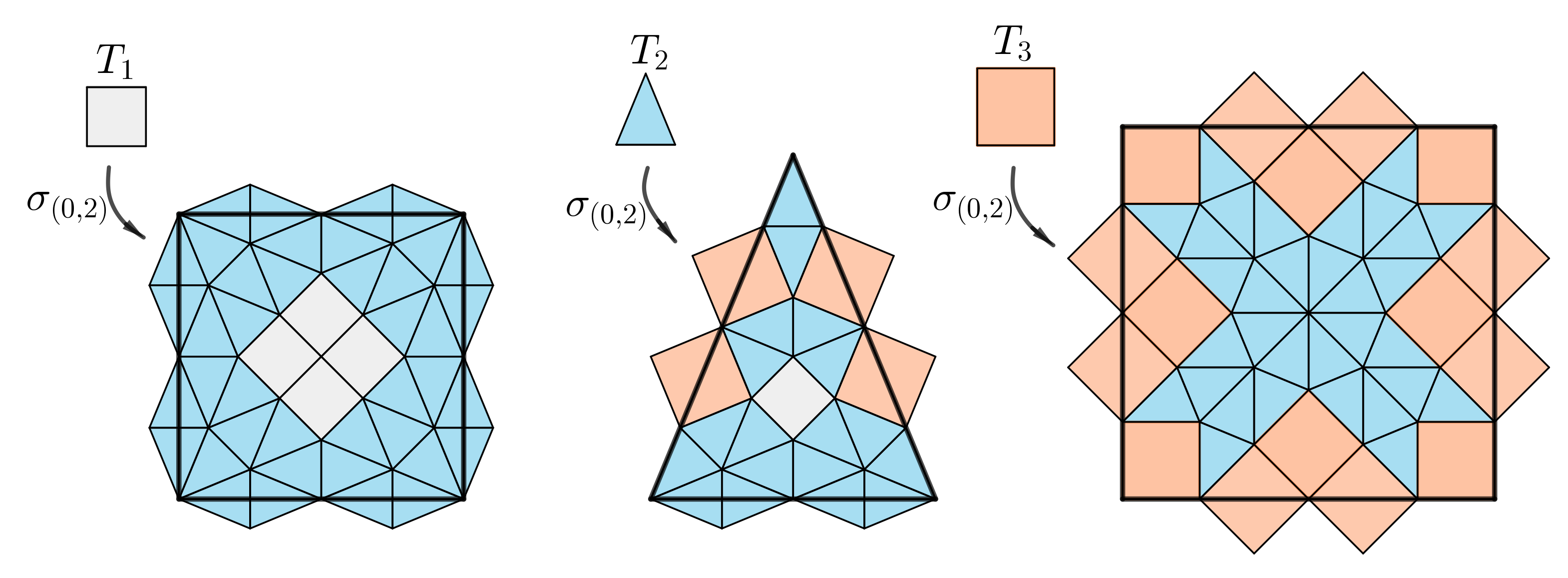}
		\caption{The substitution rule $\sigma_{(0,2)}$.}
		\label{fig:0,2substitution}
	\end{figure}
	
\begin{figure}[H]
		\centering
		\includegraphics[width=0.75\linewidth]{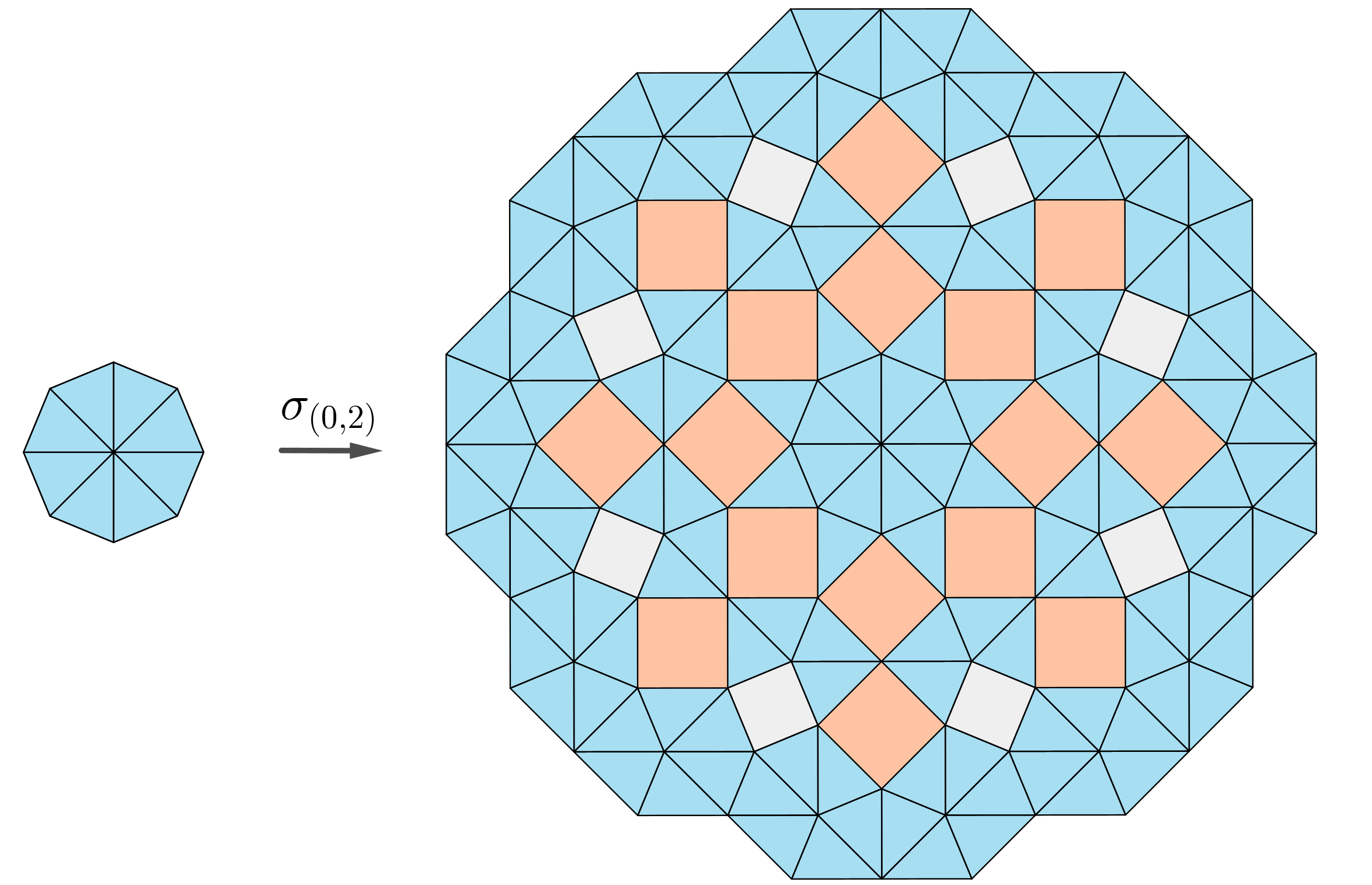}
		\caption{$\mathcal{H} \subset \sigma_{(0,2)}(\mathcal{H})$.}
		\label{fig:0,2seed}
	\end{figure}


\noindent \textbf{Case 2: $m+n$ is odd}\\

\begin{figure}[H]
		\centering
		\includegraphics[scale=.9]{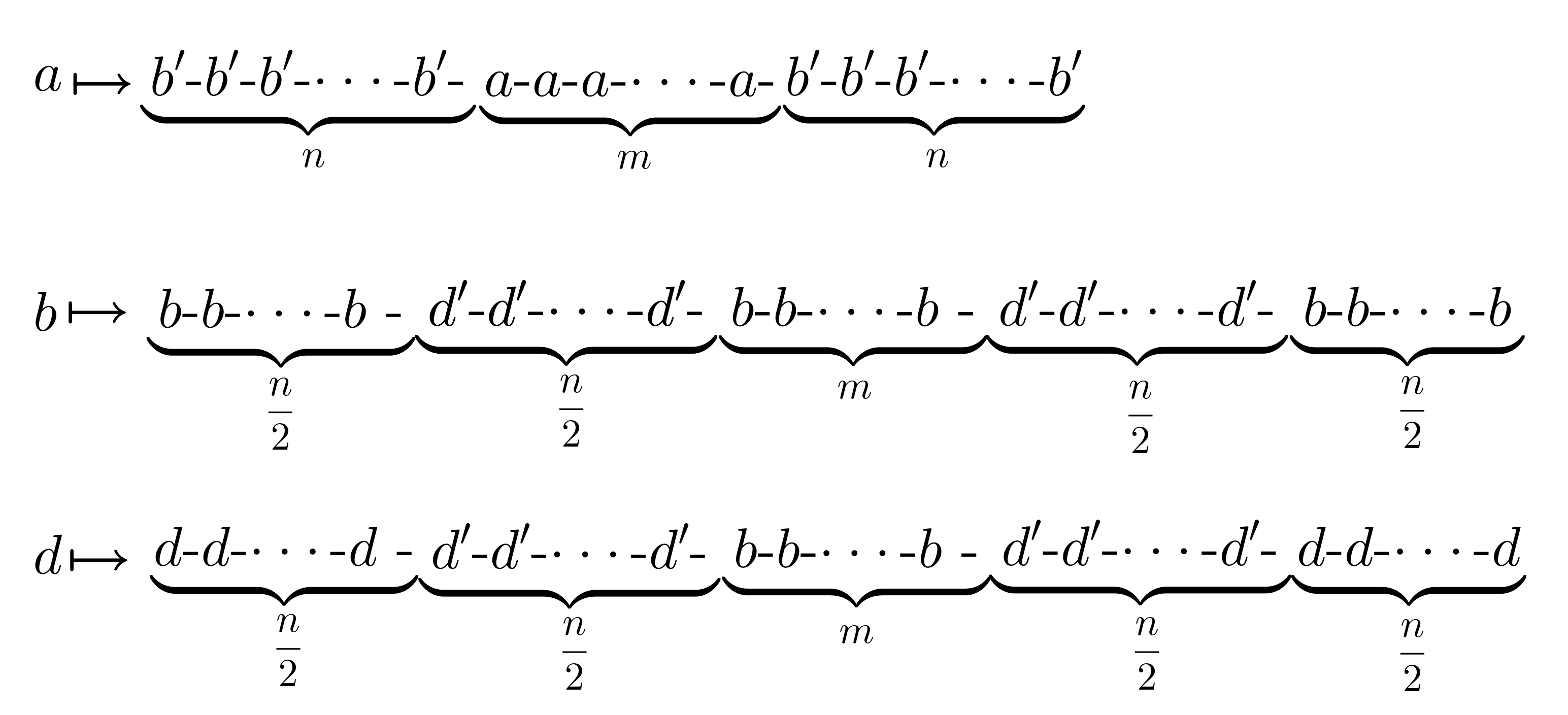}
		\caption{Edge dissection rule for the case $m+n$ is odd; the dissection for $\delta_{(m,n)}c$ is the same as that for $\delta_{(m,n)}a$.}
		\label{fig:edgesubstitutionodd}
	\end{figure}

\begin{figure}[H]
		\centering
		\includegraphics[width=.9\linewidth]{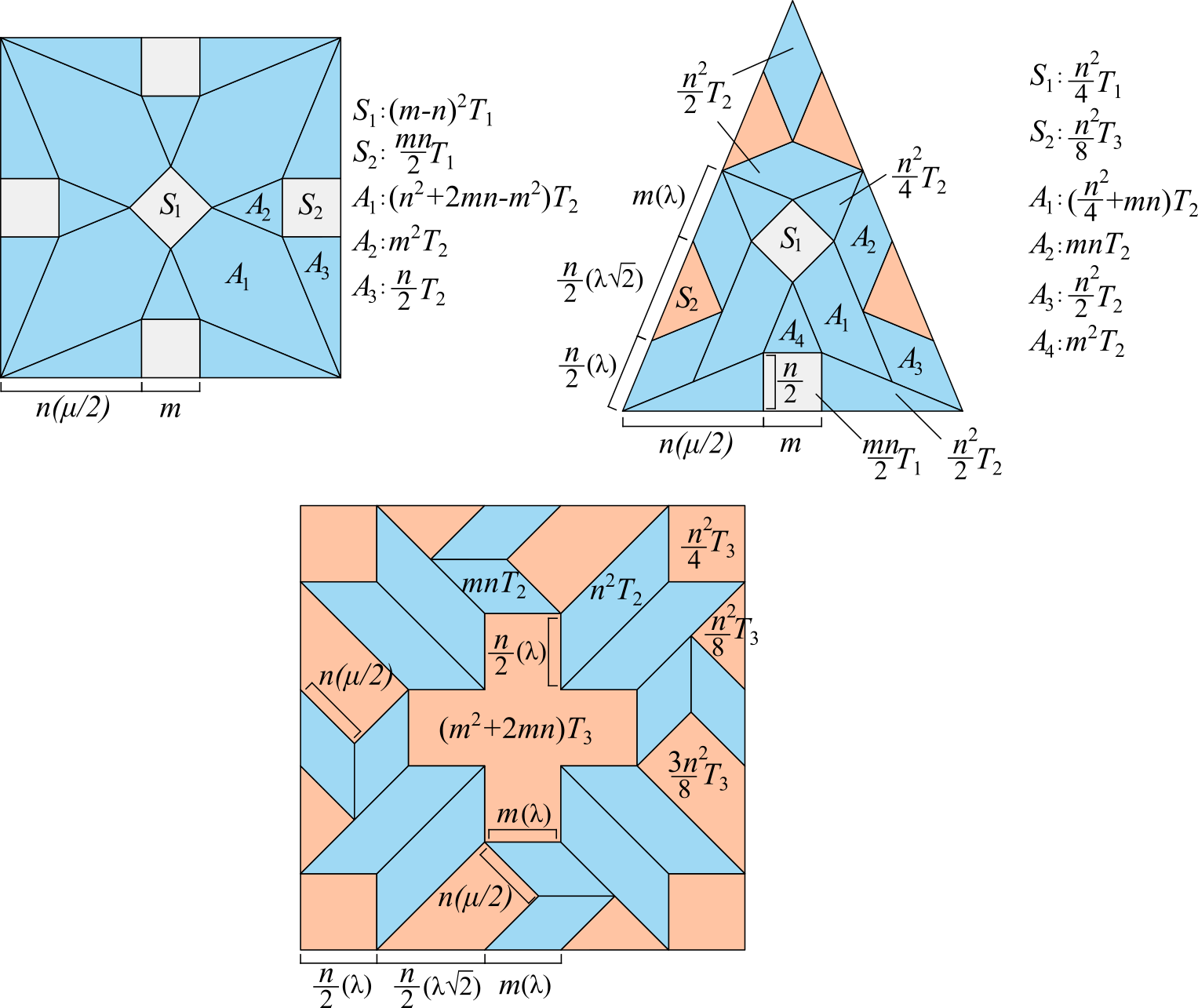}
		\caption{The dissection of $\delta_{(m,n)} T_i$, $i \in \{1,2,3\}, m<n.$}
		\label{fig:caseoddmlessn}
	\end{figure}

\begin{figure}[H]
		\centering
		\includegraphics[width=.9\linewidth]{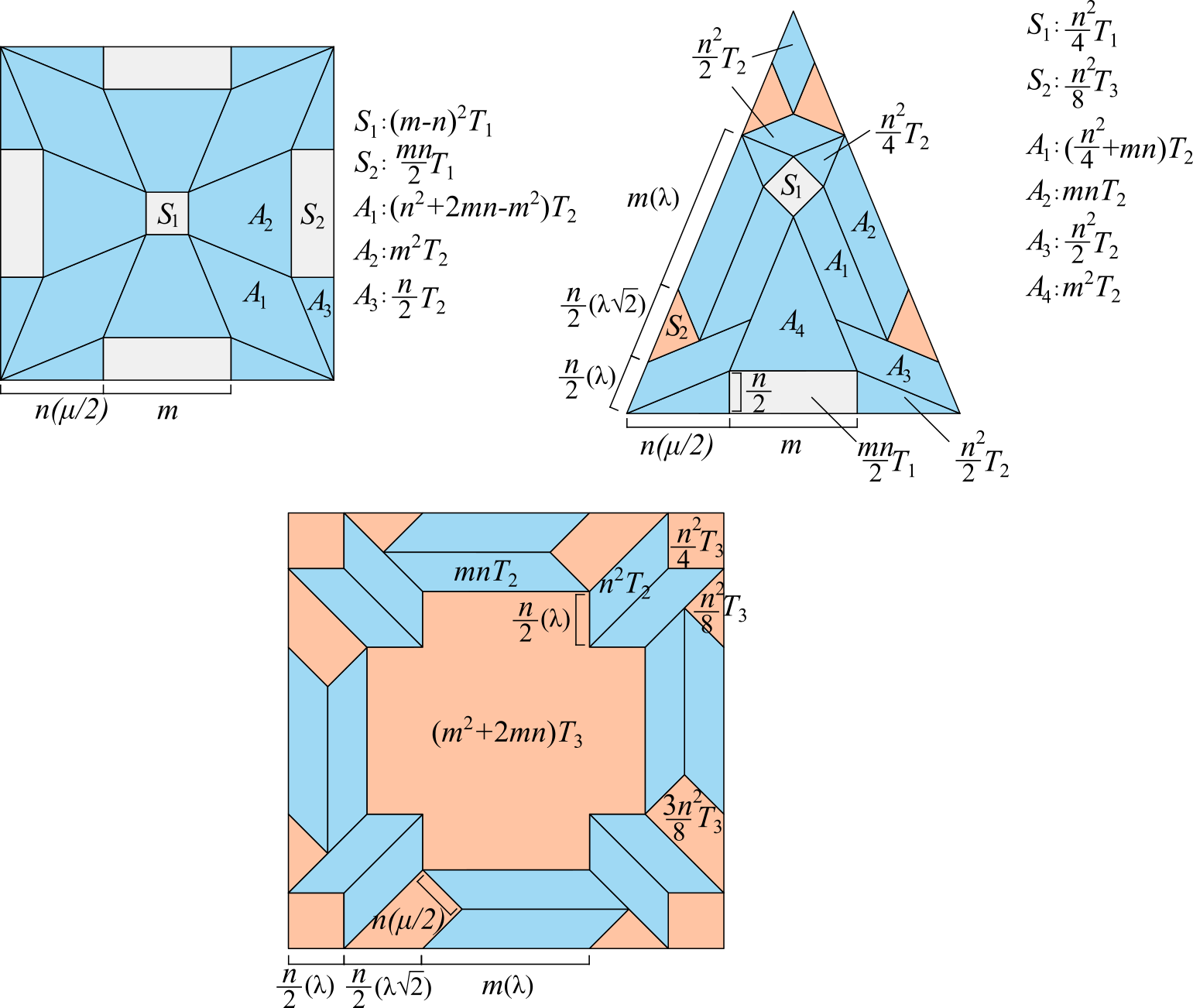}
		\caption{The dissection of $\delta_{(m,n)} T_i$, $i \in \{1,2,3\}, m>n.$}
		\label{fig:caseoddmgreatern}
	\end{figure}

\noindent Suppose now that $m+n$ is odd. The corresponding edge dissections are shown in Figure~\ref{fig:edgesubstitutionodd}. As in the even case, they determine decompositions of $\delta_{(m,n)}T_i$ into regions, each of which can be further dissected into congruent copies of a single prototile. These decompositions are shown in Figures~\ref{fig:caseoddmlessn} and~\ref{fig:caseoddmgreatern} for $m<n$
and $m>n$, respectively.

Unlike the even case, the patch $\sigma_{(m,n)}(T_3)$ is not invariant under reflection. We therefore fix the substitution patches shown in the figures and use only their rotated and translated copies throughout the construction.

\begin{figure}[H]
		\centering
		\includegraphics[width=.5\linewidth]{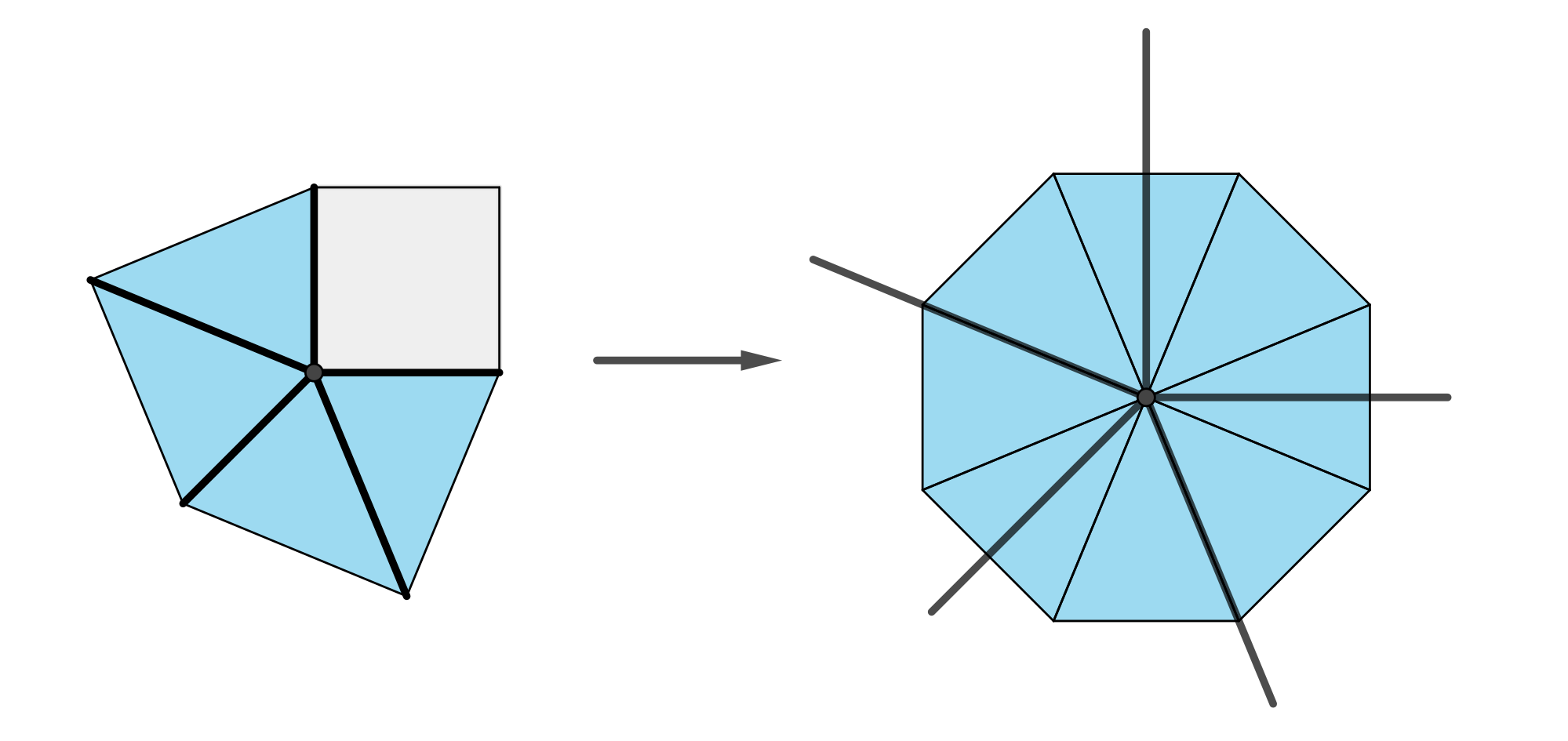}
		\caption{The patches $\mathcal{P}$ and $\mathcal{H}'$.}
		\label{fig:seedH'}
	\end{figure}

A further distinction from the even case is that the octagonal patch $\mathcal{H}$ does not occur in any first-order supertile. A congruent copy does, however, occur in the second-order supertiles of $T_1$ and $T_2$. To see this, let $\mathcal{P}$ be the patch shown in Figure~\ref{fig:seedH'}~(left), consisting of one copy of $T_1$ and several copies of $T_2$. The patch $\mathcal{P}$ occurs in the interior of both $\sigma_{(m,n)}(T_1)$ and $\sigma_{(m,n)}(T_2)$. Its image $\sigma_{(m,n)}(\mathcal{P})$ contains a patch $\mathcal{H}'$ congruent to $\mathcal{H}$, as shown in Figure~\ref{fig:seedH'}~(right). Hence, $\mathcal{H}'$ occurs in both $\sigma_{(m,n)}^2(T_1)$ and $\sigma_{(m,n)}^2(T_2)$. Since $\mathcal{H}'\subseteq\sigma_{(m,n)}(\mathcal{H}')$, the patches $\sigma_{(m,n)}^k(\mathcal{H}')$ form a nested sequence that converges to a tiling $\mathcal{T}_{(m,n)}$ corresponding to $\sigma_{(m,n)}$.

The substitution respects the rotational symmetries of each prototile and therefore preserves the rotational symmetries of $\mathcal{H}'$. Consequently, $\mathcal{T}_{(m,n)}$ is invariant under $C_8$, the cyclic group of order~$8$, and hence has $8$-fold rotational symmetry. However, because $\sigma_{(m,n)}(T_3)$ is not invariant under reflection, neither $\sigma_{(m,n)}(\mathcal{H}')$ nor $\mathcal{T}_{(m,n)}$ has reflection symmetry through the origin.

The preceding construction establishes our main theorem.

\begin{thm}\label{thm:maintheorem}
Let $m\geq 0$ and $n\geq 1$ be integers, excluding the case in which $m$ is even and $n$ is odd. Then there exists a substitution rule $\sigma_{(m,n)}$ with inflation factor $\delta_{(m,n)}=m+n\mu$ and prototile set $\{T_1,T_2,T_3\}$ that admits a tiling with $8$-fold rotational symmetry.
\end{thm}

\begin{rem}
As discussed in~\cite{saycoa2026} for the cases $(1,2)$ and $(2,2)$, variants of these dissections, and hence of the corresponding substitution rules, can be obtained by rearranging certain tiles. For $m=n$, one such variant of the dissection of $\delta_{(n,n)}T_3$ is shown in Figure~\ref{fig:variant}. At each corner of the inflated prototile, the patch consisting of the large square and the two adjacent blue regions is rotated through $\pi$. Similar variants can be constructed in the remaining cases.

	\begin{figure}[H]
		\centering
		\includegraphics[width=.5\linewidth]{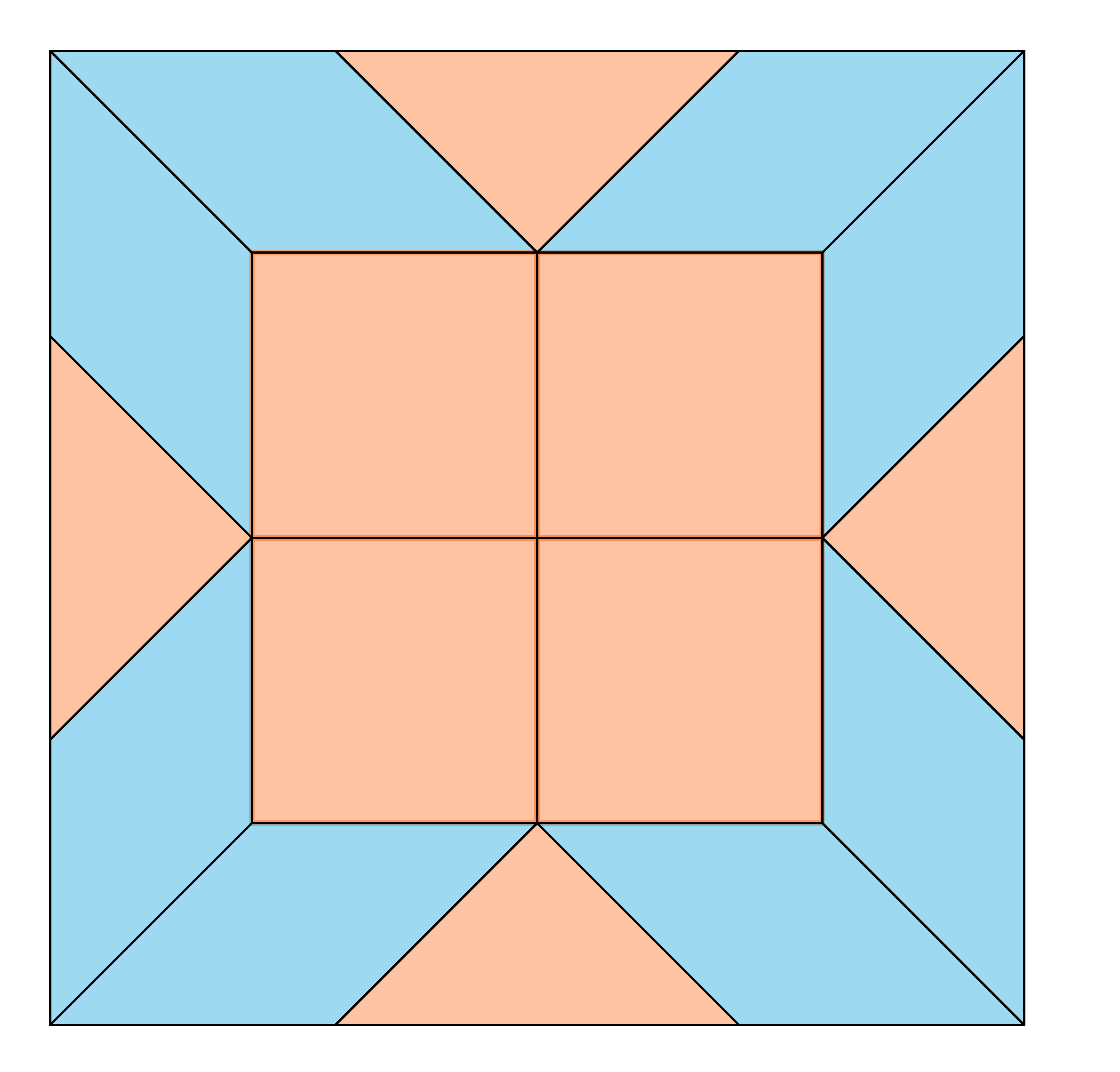}
		\caption{A variant of $\delta_{(n,n)}T_3$.}
		\label{fig:variant}
	\end{figure}
\end{rem}

\section{Basic Properties of $\sigma_{(m,n)}$}\label{sec:properties}

In addition to the existence of a tiling with $8$-fold rotational symmetry in $\mathbb{X}_{\sigma_{(m,n)}}$, several immediate properties follow from the construction.

\begin{prop}
The substitution $\sigma_{(m,n)}$ is primitive. 
\label{prop:primitive}
\end{prop}
\begin{proof}
Considering the dissections of $\delta_{(m,n)}T_i$ described in the previous section, the substitution matrix of $\sigma_{(m,n)}$ is given by 

\begin{equation}
	M_{(m,n)}=\begin{bmatrix}
		m^2 + n^2 & \dfrac{mn}{2}+b& 0 \\
		8mn + 8n^2 & m^2 + 4mn + 3n^2+4b  & 8mn + 8n^2 \\
		0 & n^2-2b & m^2 + 2mn + 3n^2
	\end{bmatrix},
	\label{eq:submatrix1}
\end{equation}

where
\[
b =
\begin{cases}
0, & \text{if } m \ge n \text{ and } m+n \text{ is even}, \\[6pt]
\dfrac{(n-m)^2}{4}, & \text{if } 0 \le m < n \text{ and } m+n \text{ is even}, \\[8pt]
\dfrac{n^2}{4}, & \text{if } m+n \text{ is odd}.
\end{cases}
\]

For each case, all entries of the matrix are positive except for the $(1,3)$- and $(3,1)$- entries. A direct computation shows that $M_{(m,n)}^2$ is strictly positive. Hence $M_{(m,n)}$ is primitive, and so is $\sigma_{(m,n)}$. 
\end{proof} 

Since $M_{(m,n)}$ is primitive, the Perron--Frobenius theorem applies, together with its standard consequences for substitution systems \cite{pytheasfogg2002,baakegrimm2013}. Its Perron--Frobenius eigenvalue is $\delta_{(m,n)}^2$, corresponding to the scaling of areas under inflation, and is the unique eigenvalue of largest modulus. The vector of prototile areas is a positive left eigenvector, while the positive right eigenvector, normalized so that its entries sum to $1$, gives the relative frequencies $\mbox{freq}(T_i)$ of the prototiles.

More precisely,
\[v=\bigl(\mbox{area}(T_1),\, \mbox{area}(T_2),\, \mbox{area}(T_3)\bigr)
=\left(1,\, \frac{1+\sqrt{2}}{4},\, \frac{2+\sqrt{2}}{2}\right)
\] and
\[
u_{(m,n)}=\bigl(\mbox{freq}(T_1),\, \mbox{freq}(T_2),\, \mbox{freq}(T_3)\bigr)^{\top}.
\]
The relative frequencies for each case are listed in Tables~\ref{tab:areafraction1} and \ref{tab:areafraction2}. The area fractions $\mbox{af}(T_i)$ are also given; they are computed using
\[
\mbox{af}(T_i)
=
\frac{\mbox{area}(T_i)\mbox{freq}(T_i)}{\sum_{j=1}^{3}\mbox{area}(T_j)\mbox{freq}(T_j)}.
\]

			\begin{table}[H]
		\begin{center}
			\begin{tabular}{| c | c | c | c |}
				\hline
				Case & Prototile  & Relative Frequency & Area Fraction\\
				\hline
				&&&\\
				$1 \leq n \leq m$ & $T_1$  & $\dfrac{m(2 - \sqrt{2})}{x_1}$ &$\dfrac{m(2 - \sqrt{2})}{y_1}$\\[6mm]
				& $T_2$ & $\dfrac{4(m+n)\sqrt{2}}{x_1}$& $\dfrac{(m + n)(2 + \sqrt{2})}{y_1}$\\[6mm]
				& $T_3$ & $\dfrac{2n}{x_1}$ & $\dfrac{n(2 + \sqrt{2})}{y_1}$\\[4mm]
				\hline 
				&&&\\
				$0 \leq m < n$ & $T_1$ &$\dfrac{(m^2+n^2)(2-\sqrt{2})}{x_2}$ &$\dfrac{(m^2+n^2)(2-\sqrt{2})}{y_2}$
				\\[6mm]
				& $T_2$ &$\dfrac{8(n^2+mn)\sqrt{2}}{x_2}$ & $\dfrac{2(n^2+mn)(2+\sqrt{2})}{y_2}$
				\\[6mm]
				& $T_3$  &$\dfrac{2(n^2+2mn-m^2)}{x_2}$  & $\dfrac{(n^2+2mn-m^2)(2+\sqrt{2})}{y_2}$
				\\[4mm]
				\hline
			\end{tabular}
			\caption{The relative frequencies and area fractions of the prototiles of $\sigma_{(m,n)}$ with $m+n$ even. Here, $x_1=2(m+n) + (3m + 4n)\sqrt{2}$, $x_2=4(n^2+mn)+(7n^2+8mn-m^2)\sqrt{2}$, $y_1=4(m+n) + 2n\sqrt{2}$, and $y_2=8(n^2+mn)-2(m^2-2mn-n^2)\sqrt{2}$.	\label{tab:areafraction1}}
			
		\end{center}
	\end{table}
	
			\begin{table}[H]
		\begin{center}
			\begin{tabular}{| c | c | c | }
				\hline
				 Prototile  & Relative Frequency & Area Fraction\\
				\hline
				&&\\
				 $T_1$ &$\dfrac{2m+n}{x_3}$  & $\dfrac{2m+n}{y_3}$
				\\[6mm]
				$T_2$  & $\dfrac{8(m+n)(1+\sqrt{2})}{x_3}$ &  $\dfrac{2(m+n)(3+2\sqrt{2})}{y_3}$
				\\[6mm]
				 $T_3$  & $\dfrac{n(2+\sqrt{2})}{x_3}$  & $\dfrac{n(3+2\sqrt{2})}{y_3}$
				\\[4mm]
				\hline
			\end{tabular}
			\caption{The relative frequencies and area fractions of the prototiles of $\sigma_{(m,n)}$ with $m+n$ odd. Here, $x_3=(10m+11n)+(8m+9n)\sqrt{2}$ and $y_3=(8m+10n)+(4m+6n)\sqrt{2}$.	\label{tab:areafraction2}}
			
		\end{center}
	\end{table}

\begin{prop}\label{prop:simple}
Every tiling corresponding to $\sigma_{(m,n)}$ is simple.
\end{prop}

\begin{proof}
The edge dissections ensure that tiles in every legal patch meet edge-to-edge. Hence, every tiling corresponding to $\sigma_{(m,n)}$ is edge-to-edge.

Let $R_\theta$ denote rotation through angle $\theta$ about the origin, and let $S$ denote reflection across a fixed line through the origin. By construction, every tile in a first-order supertile has orientation belonging to the finite dihedral group $D_{16}=\left\{R_{k\pi/16},\,R_{k\pi/16}S\mid k=0,1,\ldots,15\right\}$. The substitution preserves orientations in the sense that $\sigma_{(m,n)}(RT_i+t) = R\sigma_{(m,n)}(T_i)+\delta_{(m,n)}t$ for every allowed orientation $R$. Since $D_{16}$ is closed under composition, the orientations of the tiles in every supertile, hence in every legal patch, belong to $D_{16}$. Thus orientations of tiles in any legal patch of $\sigma_{(m,n)}$ belong to $D_{16}$. Consequently, $\mathcal{O}(\mathcal{T})\subseteq D_{16}$ for any tiling $\mathcal{T}$ corresponding to $\sigma_{(m,n)}$. Thus, $\mathcal{O}(\mathcal{T})$ is finite, and therefore $\mathcal{T}$ is simple.
\end{proof}

As an immediate consequence of the preceding propositions together with Remarks~\ref{rem:flc} and~\ref{rem:repetitive}, we obtain the following corollary.

\begin{cor}
Every tiling corresponding to $\sigma_{(m,n)}$ has FLC and is repetitive.
\end{cor}

\begin{rem}
\label{rem:dynamics} 
The preceding properties have consequences for the dynamics of the hull $\mathbb{X}_{\sigma_{(m,n)}}$, some of which we list below. See \cite{solomyak1997,baakegrimm2013,sadun2008} for further background.

\begin{enumerate}
\item The hull $\mathbb{X}_{\sigma_{(m,n)}}$ of the substitution and the translational hull of the constructed tiling $\mathcal{T}_{(m,n)}$ coincide: $\mathbb{X}_{\mathcal{T}_{(m,n)}}=\mathbb{X}_{\sigma_{(m,n)}}$. Here, $\mathbb{X}_{\mathcal{T}_{(m,n)}} =\overline{\{\mathcal{T}_{(m,n)}-x \mid x\in\mathbb{R}^2\}}$, where the closure is taken in the local topology. Intuitively, two tilings are close in this topology if, after small translations, they agree on a large neighborhood of the origin.

\item The hull $\mathbb{X}_{\sigma_{(m,n)}}$ is compact in the local topology. That is, every sequence of tilings in the hull has a subsequence that converges locally to another tiling in the hull.

\item The dynamical system $(\mathbb{X}_{\sigma_{(m,n)}},\mathbb{R}^2)$ is uniquely ergodic. Thus, every legal patch has a uniform frequency that is independent of the tiling in the hull.
\end{enumerate}
\end{rem}

As an illustration of Item~3 in Remark~\ref{rem:dynamics}, we explicitly computed the frequencies of the vertex stars and edge types for the $(1,1)$ case in~\cite{saycoa2026}.

\section{Algebraic Classification of the Inflation Factors}\label{sec:algebra}

In a cut-and-project description, points of a higher-dimensional lattice are projected into physical space, with an acceptance domain in internal space determining which projected points are selected. In~\cite{saycoa2026}, we constructed such higher-dimensional representations for the tilings associated with $(m,n)=(1,1)$ and $(1,2)$ and related their different behaviors to the PV and unimodular properties of their inflation factors. In the present work, we extend the algebraic part of this analysis to the entire family by classifying the parameter pairs according to these two properties. In standard cut-and-project and substitution settings, this classification is closely connected with contraction in internal space, the structure of acceptance domains, and the dynamical spectrum \cite{baakegrimm2013,baake2020inflation,solomyak1997}. Establishing these geometric and dynamical consequences for every member of the present family, however, requires a separate analysis.

Recall that a real algebraic integer $\alpha>1$ is called a \textit{Pisot--Vijayaraghavan number} (or simply a \textit{PV number}) if every algebraic conjugate of $\alpha$ other than $\alpha$ itself has modulus strictly less than $1$.

\begin{prop}
The inflation factor $\delta_{(m,n)}$ is a PV number if and only if
$m=\left\lfloor n(\sqrt{2}-1)\right\rfloor$ or
$m=\left\lfloor n(\sqrt{2}-1)\right\rfloor+1$.
\end{prop}

\begin{proof}
The only algebraic conjugate of $\delta_{(m,n)}$ is $\delta'_{(m,n)}=(m+n)-n\sqrt{2}=m-n(\sqrt{2}-1)$. Hence, by the definition of a PV number, $\delta_{(m,n)}$ is PV if and only if $\left|m-n(\sqrt{2}-1)\right|<1$. Since $n \geq 1$ and $n(\sqrt{2}-1)$ is not an integer, $\delta_{(m,n)}$ PV if and only if $m=\left\lfloor n(\sqrt{2}-1)\right\rfloor$ or $m=\left\lfloor n(\sqrt{2}-1)\right\rfloor+1$.
\end{proof}

\begin{figure}[H]
		\centering
		\includegraphics[width=.5\linewidth]{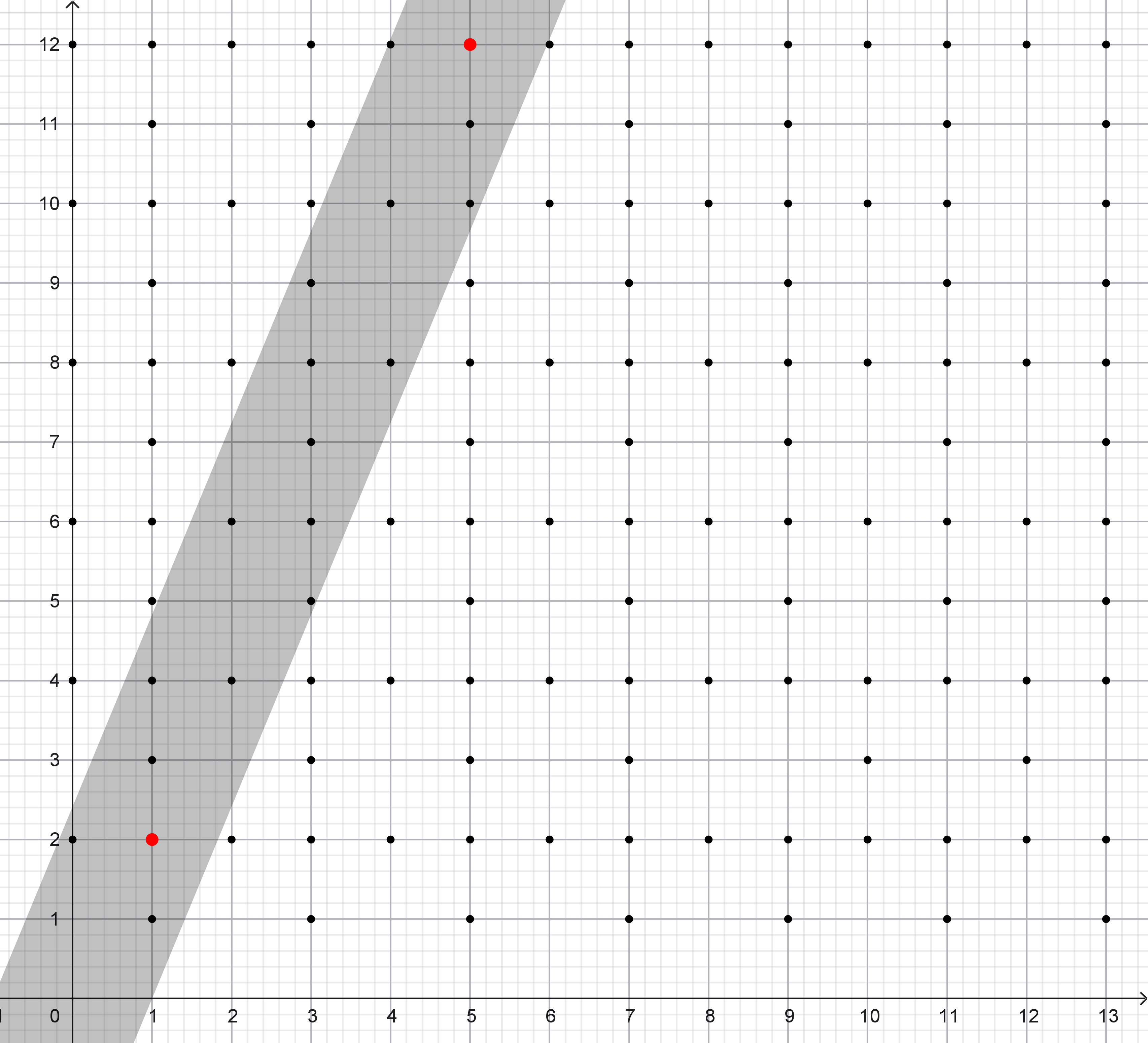}
		\caption{Admitted parameter pairs in the $(m,n)$-plane. The shaded open strip contains the pairs with PV inflation factors. Red points indicate the unit inflation factors.}
		\label{fig:strip}
	\end{figure}

Geometrically, the admitted PV parameter pairs are the integer points with $m\geq0$ and $n\geq1$ lying within the open strip $\left|m-n(\sqrt{2}-1)\right|<1$. This strip is shaded in Figure~\ref{fig:strip}, with $m$ on the horizontal axis and $n$ on the vertical axis.

A related but distinct property is whether the inflation factor is an algebraic unit. For $\alpha=a+b\sqrt{2}\in\mathbb{Z}[\sqrt{2}]$, its algebraic norm is $N(\alpha)=\alpha\alpha'=a^2-2b^2$. The element $\alpha$ is an \textit{algebraic unit} in  $\mathbb{Z}[\sqrt{2}]$ if and only if $N(\alpha)=\pm1$. For a unit inflation factor $\delta_{(m,n)}$, it follows that $|\delta_{(m,n)}'|=1/\delta_{(m,n)}<1$, since $\delta_{(m,n)}>1$. Thus, every unit inflation factor in the family is automatically a PV number. Hence, all admissible unit parameter pairs lie within the PV strip in Figure~\ref{fig:strip}, where they are indicated by the red points. In particular, there are no non-PV unit inflation factors in this family.

\begin{prop}
For the parameter pairs admitted by the construction, the inflation factor $\delta_{(m,n)}$ is an algebraic unit if and only if $\delta_{(m,n)}=(1+\sqrt{2})^{2r}$ for some $r\geq1$.
\end{prop}
\begin{proof}
Note that $N\bigl(\delta_{(m,n)}\bigr) =\delta_{(m,n)}\delta_{(m,n)}'=(m+n)^2-2n^2$. Consequently, $\delta_{(m,n)}$ is an algebraic unit if and only if $(m+n)^2-2n^2=\pm1$. Setting $x=m+n$ and $y=n$, the equation with positive sign becomes the Pell equation $x^2-2y^2=1$, whose fundamental positive solution is $(x,y)=(3,2)$, corresponding to $3+2\sqrt{2}=(1+\sqrt{2})^2$. The equation with negative sign is the negative Pell equation $x^2-2y^2=-1$, whose fundamental positive solution is $(x,y)=(1,1)$, corresponding to $1+\sqrt{2}$. Hence, all positive solutions of the two equations together are given by $x+y\sqrt{2}=(1+\sqrt{2})^k$ for $k\geq1$. Since $x=m+n$ and $y=n$, the excluded parity condition that $m$ is even and $n$ is odd is equivalent to both $x$ and $y$ being odd. For the solutions $x+y\sqrt{2}=(1+\sqrt{2})^k$, both coefficients are odd if and only if $k$ is odd. Hence, the excluded parameter pairs correspond precisely to odd $k$. Therefore, the admitted pairs require $k$ to be even. Writing $k=2r$, the admissible unit inflation factors are precisely $\delta_{(m,n)}=(1+\sqrt{2})^{2r}$ for some integer $r\geq1$.
\end{proof}

The preceding classification of the unit inflation factors has a direct interpretation in terms of unimodularity. In~\cite{saycoa2026}, we introduced the integer matrix $M^s_{(m,n)}$ describing the action of inflation on the embedding lattice $\mathcal{L}$ in superspace. The image $M^s_{(m,n)}\mathcal{L}$ is a sublattice of $\mathcal{L}$ of index 
$|\det M^s_{(m,n)}|$. The substitution is called \textit{unimodular} if $|\det M^s_{(m,n)}|=1$ and \textit{non-unimodular} if $|\det M^s_{(m,n)}|>1$.

Using the general form of $M^s_{(m,n)}$ given in~\cite{saycoa2026}, a direct calculation gives $\det M^s_{(m,n)}=(m^2+2mn-n^2)^2=N\bigl(\delta_{(m,n)}\bigr)^2$. Consequently, $\sigma_{(m,n)}$ is unimodular if and only if $\delta_{(m,n)}$ is an algebraic unit. By the preceding proposition, the admitted unimodular cases are therefore precisely those for which $\delta_{(m,n)}=(1+\sqrt{2})^{2r}$ for some integer $r\geq1$. Equivalently, let $(P_k)_{k\geq0}$ denote the sequence of Pell numbers defined by $P_0=0$, $P_1=1$, and $P_k=2P_{k-1}+P_{k-2}$ for $k\geq2$. Since $(1+\sqrt{2})^k=P_{k-1}+P_k(1+\sqrt{2})$, the admitted unimodular parameter pairs are $(m,n)=(P_{2r-1},P_{2r})$ for $r\geq1$.

This classification recovers the cases considered in~\cite{saycoa2026}: $\sigma_{(1,2)}$ is unimodular, whereas $\sigma_{(1,1)}$ is non-unimodular. The corresponding tilings were shown there to be quasiperiodic and limit-quasiperiodic, respectively. In the latter case, the perpendicular-space description involves a hierarchy of distinct windows rather than a single uniform acceptance domain, as in the theory of superquasicrystals \cite{niizeki2005superquasicrystals,fujita2006superquasicrystals}. Determining whether analogous dynamical conclusions hold throughout the present family requires a separate analysis.

\section{Perpendicular space images and Fourier spectra}\label{sec:perp}

The number of vertices created per generation in our tilings increases at an unwieldy rate with larger inflation factors, such that calculating perpendicular space images and Fourier spectra which are truly representative of the infinite tiling becomes computationally expensive. However, here we present a few images, corresponding to $\mathcal{T}_{(1,1)}$, $\mathcal{T}_{(1,2)}$, and $\mathcal{T}_{(5,12)}$. We have chosen these tilings based on their inflation factors: while all are PV numbers, $\delta_{(1,1)}$ is non-unimodular, and $\delta_{(1,2)}$, $\delta_{(5,12)}$ are unimodular. 

\begin{figure}[H]
	\centering
	\includegraphics[width=\linewidth]{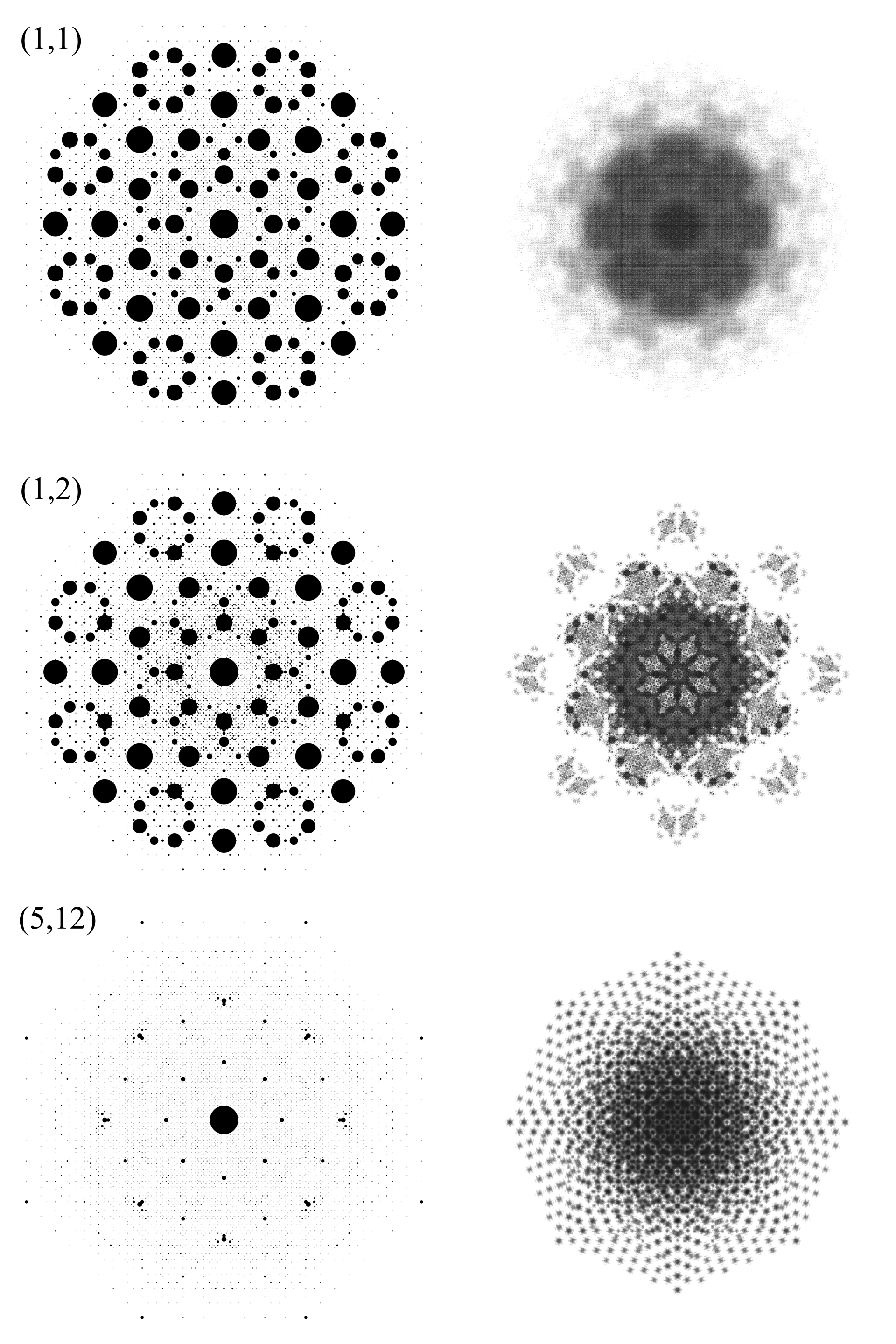}
	\caption{Left column: diffraction pattern, and right column: perp-space image of $\mathcal{T}_{1,1}$, $\mathcal{T}_{1,2}$, $\mathcal{T}_{5,12}$ respectively. }
	\label{fig:dynamics}
\end{figure}

Figure \ref{fig:dynamics} show the perpendicular space images (right) and Fourier spectra (left) of the selected tilings. The perpendicular space images are constructed by indexing each vertex in the tiling using integers $\mathbf{n} = (n_1, n_2, n_3, n_4)$ with respect to four physical space vectors ($\mathbf{e^{\parallel}_0},...,\mathbf{e^{\parallel}_3}$), where $\mathbf{e^{\perp}} = (\mathbf{e^{\perp}_0,...,\mathbf{e^{\perp}_3}})$ are the associated perpendicular space vectors. The image is then computed by $\mathbf{n}\cdot \mathbf{e^{\perp}}$. The Fourier spectra are calculated using the superspace formalism  \cite{elser1985indexing, elser1986diffraction}: reciprocal lattice vectors are enumerated as integer linear combinations $G = K\cdot m$, where K is the reciprocal basis matrix and $m \in [-5, 5]^4$. Each vector is then decomposed into parallel- and perpendicular-space components $r^\parallel$ and $r^\perp$, where diffracted intensities were evaluated as $|\sigma_j e^{-2\pi G^{\perp}\cdot r^\perp}|^2$  and plotted at their parallel-space positions. Here, the radius of each circle corresponds to the calculated intensity, normalised by the maximum at the origin.

%
%

\section{Comparison to experimental work}\label{sec:comparison}

As discussed in the introduction, it was shown by Fayen et al. that different octagonal phases of hard sphere systems can be formed by changing the ratio of small to large spheres \cite{fayen2023self}. For a given size ratio, the fraction of small spheres $x_s$ changes the tile area fractions observed in the resultant phases. In other words, tuning $x_s$ allows the system to sample different (randomised) octagonal tilings. Here we show that our parameters $(m.n)$ allow us to create families of tilings which exactly map onto the observed experimental tile area fractions, such that we can say that the observed phases can be expressly modelled as randomised versions of our tilings.

We start by taking the formulae which give the tile area fractions in terms of $x_s$ from \cite{fayen2023self}. Then, we rearrange for $x_s$:

\begin{align}
A_1 = \dfrac{-(4+\sqrt{2})x_s+4}{6-4x_s} \Longrightarrow x_s& = \dfrac{6A_1-4}{4A_1-\sqrt{2}-4},\\
A_2 = \dfrac{-(8+5\sqrt{2})x_s+4\sqrt{2}+7}{6-4x_s} \Longrightarrow x_s& = \dfrac{6A_2-4\sqrt{2}-7}{4A_2-5\sqrt{2}-8},\\
A_3 = \dfrac{2(4+3\sqrt{2})x_s-4\sqrt{2}-5}{6-4x_s}\Longrightarrow x_s& = \dfrac{6A_3+4\sqrt{2}+5}{4A_3+6\sqrt{2}+8},
\end{align}
such that we can extract the minimum and maximum values for $x_s$, determined only by our tiling parameters, obtained by substituting any expression for tile area fractions into these relationships. For example, doing so for case $1\leq n \leq m$, and arbitrarily using $A_1$ gives

\begin{equation}
x_s = \dfrac{6m(2-\sqrt{2}-4y_1)}{4m(2-\sqrt{2})-y_1(\sqrt{2}-4)}.
\label{eq:case1xs}
\end{equation}

More generally, the limits of $x_s$ in these relationships are determined only by the denominators of the area fractions i.e., $y_1, y_2, y_3$ from Tables \ref{tab:areafraction1}, \ref{tab:areafraction2}, considering the constraints of the relevant cases. For instance, Equation \ref{eq:case1xs} is minimal when $y_1$ is maximal, or when $n=1$ and $m \rightarrow \infty$, such that $x_s \rightarrow \frac{1}{4}(4-\sqrt{2}) \approx 0.646...$. By inspecting each expression in a similar fashion, we summarise the limits for $x_s$ and the corresponding tile area fractions in Table \ref{tab:areafractionxs}, noting that these limits hold whether $m+n$ is even or odd.

			\begin{table}[htbp]
	\begin{center}
		\begin{tabular}{| c | c | c |}
			\hline
			Case & $x_s$ limit  & Area Fractions\\
			\hline
			&&\\
			$1 \leq n \leq m$ & min = $ \dfrac{4-\sqrt{2}}{4}$  &$A_1 = \dfrac{2-\sqrt{2}}{4}, A_2 = \dfrac{2+\sqrt{2}}{4}, A_3 = 0 $\\[6mm]
			& max = $\dfrac{7-2\sqrt{2}}{6}$& $A_1 = \dfrac{2-\sqrt{2}}{8}, A_2 = \dfrac{4+\sqrt{2}}{8}, A_3 = \dfrac{1}{4}$\\[6mm]
			\hline 
			
			&&\\
			$0 \leq m < n$ & min = $\dfrac{\sqrt{2}}{2}$ &$A_1 = \dfrac{5-3\sqrt{2}}{14}, A_2 = \dfrac{3+\sqrt{2}}{7}, A_3 = \dfrac{3+\sqrt{2}}{14}$
			\\[6mm]
			& max = $\dfrac{37-9\sqrt{2}}{34}$ & $A_1 = \dfrac{3-2\sqrt{2}}{4} , A_2 =  \dfrac{1+\sqrt{2}}{4} , A_3 = \dfrac{\sqrt{2}}{4}$
			\\[6mm]

			\hline
		\end{tabular}
		\caption{The analytically derived minimum and maximum values of the small sphere ratio $x_s$ for the octagonal phases discovered in \cite{fayen2023self}, alongside their corresponding tile area fractions.	\label{tab:areafractionxs}}
		
	\end{center}
\end{table}

\section{Conclusions and outlook}\label{sec:conclusion}

We have constructed an infinite two-parameter family of substitution rules $\sigma_{(m,n)}$ on three prototiles. The inflation factors are $\delta_{(m,n)}=m+n\mu$ for all integers $m\geq0$ and $n\geq1$, except when $m$ is even and $n$ is odd. The family contains the four examples of~\cite{saycoa2026} as the cases $(1,1)$, $(1,2)$, $(2,2)$ and $(3,2)$. It extends the silver-mean inflation of the Ammann--Beenker tiling to an infinite set of inflation factors in $\mathbb{Z}[\sqrt{2}]$. For every admissible pair, $\sigma_{(m,n)}$ admits a tiling with $8$-fold rotational symmetry, with full $D_8$ symmetry when $m+n$ is even and chiral $C_8$ symmetry when $m+n$ is odd. All substitutions in the family are primitive, and all associated tilings are edge-to-edge with finitely many tile orientations. Hence they have finite local complexity and are repetitive. The prototile frequencies and area fractions are available in closed form for every member.

On the algebraic side, $\delta_{(m,n)}$ is a PV number precisely for the pairs lying in the strip $|m-n(\sqrt{2}-1)|<1$. The substitution is unimodular precisely when $\delta_{(m,n)}=\mu^{2r}$ for some $r\geq1$, that is, for $(m,n)=(P_{2r-1},P_{2r})$. The unimodular members are therefore increasingly sparse within the PV strip, while all admissible pairs outside the strip yield non-PV inflation factors. The perpendicular-space images and Fourier spectra computed for $\mathcal{T}_{(1,1)}$, $\mathcal{T}_{(1,2)}$ and $\mathcal{T}_{(5,12)}$ illustrate the difference between the non-unimodular and unimodular PV cases.

The family provides a structural framework for the octagonal random-tiling quasicrystals observed in binary hard spheres on a plane~\cite{fayen2023self} and in vibrated granular mixtures~\cite{plati2024quasi}, which are built from exactly the same three tiles and were the initial motivation for our construction. Trivially, the area fractions of every member of the family satisfy the linear constraint required for $8$-fold symmetry~\cite{fayen2023self}. As $(m,n)$ varies, the corresponding fractions of small spheres are dense in the interval $(4-\sqrt{2})/4<x_s<(37-9\sqrt{2})/34$, that is, approximately $0.646<x_s<0.714$. This interval contains the composition $x_s=0.675$ at which the octagonal phase was imaged in both simulation and experiment: the self-assembled phases can thus be regarded as randomised counterparts of the deterministic tilings in our family. Compositions between the upper end of this interval and the Ammann--Beenker value $x_s=(8-2\sqrt{2})/7\approx0.739$ are not reached, because the area fraction of small squares in every member exceeds $(3-2\sqrt{2})/4$. Covering this window would require different constructions in which the proportion of $T_1$ can be made arbitrarily small.

Several directions remain open. First, since the self-assembled phases are random tilings, a natural next step is to randomise the present constructions. This could be achieved through random substitutions that choose independently among variant decompositions, such as Figure~\ref{fig:variant}. Alternatively, phason-flip Monte Carlo sampling could be initialised at $\mathcal{T}_{(m,n)}$ at fixed composition. Comparing the resulting diffraction patterns and perpendicular-space distributions with simulated and experimental configurations would test how closely the observed phases follow their deterministic parents. 

Second, decorating the tiles of $\mathcal{T}_{(m,n)}$ with large and small spheres as in~\cite{fayen2023self} gives explicit initial configurations for particle simulations. These could be used to compare the thermodynamic stability of the deterministic tilings with that of their randomised counterparts. 

Third, as $m$ and $n$ grow, the supertiles contain increasingly large regions tiled by a single prototile. These locally periodic domains essentially act as crystallites embedded or arranged in a quasiperiodic matrix. Quantifying their influence on the diffraction pattern is an interesting problem, as is their possible relation to the coexisting domains and defects seen in self-assembly.

Fourth, the spectral and dynamical consequences of the PV and unimodular classification should be established for the whole family. This includes the construction of acceptance domains for the unimodular members, the window hierarchies expected in the non-unimodular PV cases, and the nature of the spectrum for non-PV inflation factors~\cite{solomyak1997, baake2020inflation}.

Finally, the parity class with $m$ even and $n$ odd is excluded here by the requirement of symmetric edge dissections, but can be accessible through asymmetric dissections. Further work will demonstrate how these substitutions also admit tilings.

\section*{Acknowledgments}

SC acknowledges support fromEPSRC grant EP/X011984/1.

\section*{Conflict of interest}
\noindent The authors have no conflicts of interest to declare.

\bibliographystyle{plain}
\bibliography{references}
\end{document}